\documentclass[preprint,prd,tightenlines,nofootinbib]{revtex4}
\usepackage{amsmath,bm}
\usepackage{graphicx}
\def\mi{{\mathrm i}}
\newbox\charbox
\newbox\slabox
\def\s#1{{      
        \setbox\charbox=\hbox{$#1$}
        \setbox\slabox=\hbox{$/$}
        \dimen\charbox=\ht\slabox
        \advance\dimen\charbox by -\dp\slabox
        \advance\dimen\charbox by -\ht\charbox
        \advance\dimen\charbox by \dp\charbox
        \divide\dimen\charbox by 2
        \raise-\dimen\charbox\hbox to \wd\charbox{\hss/\hss}
        \llap{$#1$}
}}

\begin{document}

\title{Numerical integration of one-loop 
Feynman diagrams for $N$-photon amplitudes }

\author{Zolt\'an Nagy}
\affiliation{
Physics Department,
Theoretical Group,
CERN,
CH 1211 Geneva 23, Switzerland}
\author{Davison E.\ Soper}
\affiliation{
Institute of Theoretical Science,
University of Oregon,
Eugene, OR  97403-5203, USA
}

\preprint{CERN-PH-TH/2006-193}

\begin{abstract}
In the calculation of cross sections for infrared-safe observables in high energy collisions at next-to-leading order, one approach is to perform all of the integrations, including the virtual loop integration, numerically. One would use a subtraction scheme that removes infrared and collinear divergences from the integrand in a style similar to that used for real emission graphs. Then one would perform the loop integration by Monte Carlo integration along with the integrations over final state momenta. In this paper, we explore how one can perform the numerical integration. We study the $N$-photon scattering amplitude with a massless electron loop in order to have a case with a singular integrand that is not, however, so singular as to require the subtractions. We report results for $N = 4$, $N = 5$ with left-handed couplings, and $N=6$.
\end{abstract}

\date{10 November 2006}

\maketitle


\section{Introduction}
\label{sec:introduction}

The calculation of cross sections in the Standard Model and its extensions at next-to-leading order (NLO) in perturbation theory inevitably involves computing virtual loop Feynman diagrams. The standard method for this involves computing the loop integrals analytically. Once the one loop amplitude is known analytically, the result can be inserted into a calculation of the cross section in which integrals over the momenta of final state particles are performed numerically. This is the method that was introduced in Ref.~\cite{ERT} and is used, for example, in the packages {\tt MCFM} \cite{MCFM} and {\tt NLOJet++} \cite{NLOJet}.

This approach is powerful and has been successfully applied to a number of processes of experimental interest. There has been considerable progress \cite{progress} in expanding the range of processes for which an analytical answer is known.\footnote{Here ``known'' may mean that there exists a computer program to calculate the desired scattering amplitude in terms of known master integrals or other special functions. There are many approaches, some of which are conventionally called ``semi-numerical'' because parts of the calculation involve a numerical approach.} One may hope that the analytical approach may develop into a completely automatic way of generating scattering amplitudes for a wide class of processes. However, the complexity of the results produced by known analytical methods grows rapidly with the number of partons involved in the scattering. For this reason, there may be limits to the range of processes for which analytical methods are useful.

One wonders whether a wider range of processes might be amenable to calculation if one were, instead, to use numerical integration for the virtual loop integrals. In a calculation of a cross section, the numerical integration would be performed along with the integrations over the momenta of final state particles, so that there would be a single integration over a large number of variables, with the integration performed by Monte Carlo style numerical integration. The purely numerical approach will inevitably have its limitations, just as the analytical approach does. However the nature of the limitations will be different. For this reason, we believe that one should try to develop the numerical method as far as possible and see how far back the limitations can be pushed. Eventually, this should involve trying several variations on the basic theme of performing the integrations numerically. 

There are already some methods available for doing the virtual loop integrals numerically. In one method \cite{beowulfPRL,beowulfPRD}, one performs the integral over the energy flowing around the loop analytically by closing the integration contour in the upper or lower half plane and evaluating the residues of the poles in the complex energy plane that come from the propagator denominators. This is a purely algebraic step. Then the integral over the space momentum is performed numerically. There are infrared divergences, but these cancel inside the integrals between real and virtual graphs that make up a NLO cross section. This method has been applied to $e^+e^- \to 3\ {\rm jets}$ and is completely practical in that application. For more complicated processes, we do not know how to arrange a calculation in this style without subtractions. One could add subtractions to the method of Refs.~\cite{beowulfPRL,beowulfPRD}, but for this paper we have chosen a different approach.

Another method \cite{Binoth} involves transforming the loop integral into the standard Feynman parameter representation that one uses for analytically evaluating such integrals. Then the integral over the Feynman parameters is to be performed numerically. This method shows promise, but is limited by the complexity introduced by expanding the numerator functions involved. The method introduced in this paper makes use of the Feynman parameter representation while avoiding the complexities introduced by the numerator function.

This paper represents the second step of a program for calculating virtual loop integrals numerically. In the first step \cite{NSsubtractions}, we attacked the problem of infrared divergences. Typically, the integrals that one wants to evaluate have infrared divergences associated with the momenta of particles in the loop becoming collinear with the momenta of massless external particles or becoming soft. In Ref.~\cite{NSsubtractions}, we proposed a subtraction scheme in which one subtracts certain counter terms from the integrand, then adds these same counter terms back. After summing over graphs and performing the integrals analytically, the counter terms that we added back have a simple form that is easily included in the calculation of a cross section. Meanwhile, the main integrand minus the counter term integrands combine to make an integrand that is free of singularities strong enough to make the integral divergent. Thus one can numerically integrate the main integrand minus the counter-term integrands.

Despite the beauty of this approach, it is one thing to say that one can numerically integrate the combined integrand and it is another thing to do it. One needs a practical method for doing it. That is what we propose in this paper.

In order to keep our discussion reasonably simple, we attack a simple problem in which the counter terms are not present because the original integral is infrared finite. The problem is to compute the amplitude in quantum electrodynamics for scattering of two photons to produce $N-2$ photons by means of an electron loop. Our formulas include the possibility of a non-zero electron mass, but in order to face up to the problem of infrared singularities that appear when the electron mass vanishes, we concentrate on the mass zero case.

\begin{figure}
\centerline{
\includegraphics[width = 6 cm]{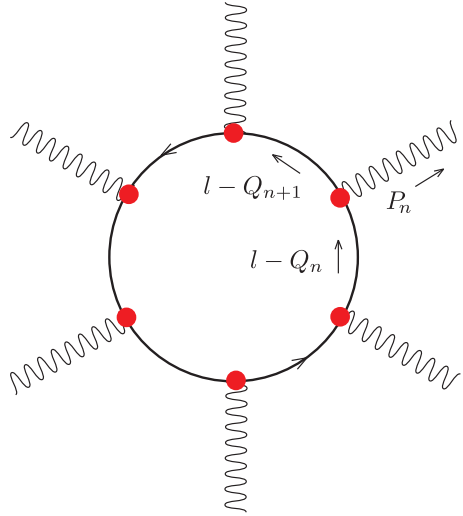}
}
\medskip
\caption{
Feynman diagram for the $N$-photon amplitude.
}
\label{fig:Nphoton}
\end{figure}

The process is illustrated in Fig.~\ref{fig:Nphoton}. Electron line $n$ in the loop carries momentum $l - Q_n$, where $Q_n$ is fixed and we integrate over $l$. The momentum carried out of the graph by external photon $n$ is\footnote{Throughout this paper, we adopt a cyclic notation for indices in the range $\{1,2,\cdots,N\}$. Thus Eq.~(\ref{eq:Qndef}) for $n = N$ is $P_N = Q_1 - Q_N$.} 
\begin{equation}
P_n = Q_{n+1} - Q_n \;\;,
\label{eq:Qndef}
\end{equation}
with $P_n^2 = 0$. The propagator denominators provide factors that would lead to logarithmic divergences after integration over the soft and collinear regions. However, these divergences are cancelled. For each electron line there is a factor $(\s l - \s Q_n)$. Thus the numerator provides a factor that removes the soft divergence from the integration region $(l - Q_n)\to 0$. Similarly at each vertex there is a factor $(\s l - \s Q_{n+1})\ \s\epsilon_n(P_n)\ (\s l - \s Q_{n})$, where $\epsilon_n(P_n)$ is the polarization vector of the photon. In the collinear limit $(l - Q_n) \to x P$, this gives a factor  $-x(1-x) \s P_n \s\epsilon_n(P_n) \s P_n  = - 2x(1-x) \s P_n\ \epsilon_n(P_n)\cdot P_n$. This vanishes because $\epsilon_n(P_n)\cdot P_n = 0$. Thus the numerator also provides a factor that removes each collinear divergence. The loop integral is also finite in the ultraviolet as long as $N>4$. (For $N=4$ the integral is divergent by power counting, so a special treatment, discussed later in this paper, is needed.) Thus we can present an algorithm that is uncluttered by the counter terms by means of using the scattering of two photons to produce $N-2$ photons. We reserve the full case of massless quantum chromodynamics for a future paper.

\section{The amplitude}
\label{sec:amplitude}

We wish to calculate the amplitude for scattering of two photons to produce $N-2$ photons by means of a (massless) electron loop. However, we formulate the problem in a more general fashion. The amplitude for any one loop graph can be represented as
\begin{equation}
{\cal M} = \int\! \frac{d^4 l}{(2\pi)^4}\ e^N N(l)
\prod_{i=1}^N\frac{1}{ (l - Q_i)^2 - m_i^2 + \mi 0}
\;\;.
\label{eq:lspace0}
\end{equation}
Here there is a loop with $N$ propagators as illustrated in Fig.~\ref{fig:Nphoton}. The $n$th propagator carries momentum $l - Q_n$ and represents a particle with mass $m_n$. At the $n$th vertex, momentum $P_n = Q_{n+1} - Q_n$ leaves the graph. In the case to be considered, all the $m_n$ and $P_n^2$ vanish, but we leave the masses and external leg virtualities open in the general formulas. There is a coupling $e$ for each vertex, where $e$ is the charge of the fermion. There is a numerator factor that, for the photon scattering case with zero electron mass, has the form
\begin{equation}
N(l) = 
{\rm Tr}\left\{
\s\epsilon_N(P_N)\ (\s l - \s Q_{N})\cdots
\s\epsilon_1(P_1)\ (\s l - \s Q_{1})
\right\}
\;\;,
\label{eq:numeratordef}
\end{equation}
where $\epsilon_i(P_i)$ is the polarization vector of photon $i$ and $e$ is the electromagnetic coupling.\footnote{Specifically, $\epsilon_i(P_i)$ for an outgoing photon is $\epsilon^*(P_i,s_i) = \epsilon(P_i,-s_i)$, where $s_i$ is the helicity of the photon. For an incoming photon, we follow the convention of using a helicity label $s_i$ equal to the negative of the physical helicity of the photon. Then $\epsilon_i(P_i)$ is $\epsilon(-P_i,-s_i)$.} In other examples, one would have a different numerator function. The only property that we really need is that $N(l)$ is a polynomial in $l$.  

It will prove convenient to modify this by inserting factors $\mi m_0^2$ in the numerator and the denominator, where $m_0^2$ is an arbitrary parameter that we can take to be of the order of a typical dot product $Q_i\cdot Q_j$. This
factor is not absolutely needed for the purposes of this paper, but it is quite useful in the case of the subtraction terms to be considered in future papers and is at least mildly helpful in the analysis of this paper. With this extra factor, we write
\begin{equation}
{\cal M} = \int\! \frac{d^4 l}{(2\pi)^4}\ \frac{\mi m_0^2\, e^N N(l)}{\mi m_0^2}
\prod_{i=1}^N\frac{1}{ (l - Q_i)^2 - m_i^2 + \mi 0}
\;\;.
\label{eq:lspace}
\end{equation}

\section{Representation with Feynman parameters}
\label{sec:feynman}

In principle, it is possible to perform the integration represented in Eq.~(\ref{eq:lspace}) directly by Monte Carlo numerical integration on a suitably deformed integration contour. We have looked into this and conclude that it may be a practical method. However, one has to pay attention to the singularities on the surfaces $(l - Q_i)^2 = m_i^2$. The geometry of these surfaces and of their intersections is somewhat complicated. There is a standard method for simplifying the singularity structure: changing to a Feynman parameter representation. One way of using this method has been emphasized in the context of numerical integrations in Ref.~\cite{Binoth}. It is the Feynman parameter method that we explore in this paper.

The Feynman parameter representation of Eq.~(\ref{eq:lspace}) is
\begin{equation}
\label{eq:Feynman1}
{\cal M} = \Gamma(N+1)
\int_0^1\! dx^0\int_0^1\! dx^1\cdots \int _0^1\! dx^N\
\delta\!\left(\sum_{i=0}^N x^i - 1\right)\
\int\! \frac{d^4 l}{(2\pi)^4}\
\frac{\mi m_0^2\,e^N N(l)}{ [D(l)+ \mi 0]^{N+1}}
\;\;.
\end{equation}
The denominator here is
\begin{equation}
D(l) = 
\sum_{i=1}^N x^i [(l - Q_i)^2 - m_i^2 ]
+\mi x^0 m_0^2
\;\;.
\end{equation}
The denominator comes with a ``$+\mi 0$'' prescription for avoiding the possibility that the integrand has a pole on the integration contour and the $+\mi x^0 m_0^2$ term serves the same purpose. With repeated use of $\sum_{i=1}^N x^i = 1 - x^0$, the denominator can be simplified to
\begin{equation}
D(l) = \frac{1}{1-x^0}
\left\{
\tilde l^2 + \Lambda^2(x)
\right\}
\;\;.
\end{equation}
Here
\begin{equation}
\tilde l = \sum_{i=1}^N x^i (l -  Q_i)
\label{eq:tildeldef}
\end{equation}
and 
\begin{equation}
\Lambda^2(x) = \frac{1}{2} \sum_{i,j = 1}^N x^i x^j S_{ij}
+ \mi \sum_{j=1}^N x^0 x^j m_0^2
\;\;,
\label{eq:Lambdasqdef}
\end{equation}
where we have defined 
\begin{equation}
\label{eq:Sijdef}
\begin{split}
S_{ij} ={}&
(Q_i - Q_j)^2 -  m_i^2 - m_j^2
\;\;.
\end{split}
\end{equation}

We change integration variables from $l$ to $\tilde l$ as given in Eq.~(\ref{eq:tildeldef}). The inverse relation is
\begin{equation}
\label{eq:tildelinverse}
l = l(\tilde l,x) 
\equiv
\frac{1}{1-x^0}\,\left(\tilde l + \sum_{i = 1}^N x^i Q_i\right)
\;\;.
\end{equation}
With these results, we have
\begin{equation}
\label{eq:lxspace}
\begin{split}
{\cal M} ={}&
\mi m_0^2 e^N\Gamma(N+1) 
\int_0^1\! dx^1\cdots \int _0^1\! dx^N\
	\theta\!\left(\sum_{i=1}^N x^i < 1\right)
\left(\sum_{i=1}^N x^i\right)^{N-3}
\\
&	\times	\int\! \frac{d^4 \tilde l}{(2\pi)^4}\
\frac{ N\!\left(l(\tilde l,x)\right)}
	{ \left[\tilde l^2 + \Lambda^{\!2}(x) + \mi 0\right]^{N+1}}
\;\;.
\end{split}
\end{equation}
Here we understand that the Feynman parameter $x^0$ is given by $x^0 = 1-
\sum_{i=1}^N x^i$.

If we wished to perform the integration analytically, the next step would be carry out the integration over $\tilde l$. However for a numerical integration, such a step would be a step in the wrong direction. Performing the $\tilde l$ integration analytically would require expanding the complicated numerator function in powers of $\tilde l$. For this reason, we leave the $\tilde l$ integration to be carried out numerically after a little simplification.

The simplification is to change variables from $\tilde l$ to a momentum $\ell$ that has been scaled by a factor $\Lambda(x)$ and rotated in the complex plane:
\begin{equation}
\begin{split}
\tilde l^{\mu}(x,\ell) ={}& 
{\textstyle{\frac{1}{2}}}
\Lambda(x)
\left\{
(1 - \mi)\ell^{ \mu}
+  
(1 + \mi)P^\mu_\nu \ell^{\nu}
\right\}
\;\;.
\label{eq:elldef}
\end{split}
\end{equation}
Here $\hat \ell^\mu = P^\mu_\nu \ell^{\nu}$ is the parity transform of $\ell$:
$\hat \ell^0 = \ell^0$, $\hat \ell^j = - \ell^j$ for $j \in \{1,2,3\}$. We have defined $\Lambda(x)$ for real $x$ and $m_0^2 \to 0$ to be $\sqrt{\Lambda^2(x)}$ if $\Lambda^2(x)$ is positive and $\mi\sqrt{-\Lambda^2(x)}$ if $\Lambda^2(x)$ is negative. The square of $\tilde l$ is
\begin{equation}
\begin{split}
\tilde l^2 
={}&
\Lambda^2(x)\,
\ell^{\mu} P_{\mu\nu}\ell^{\nu}
\;\;.
\end{split}
\end{equation}
Note that $\ell^{\mu} P_{\mu\nu}\ell^{\nu}$ is the square of $\ell$ with a euclidian inner product and is thus strictly positive.

Our integral now is
\begin{equation}
\begin{split}
{\cal M}  ={}& 
- m_0^2 e^N \Gamma(N+1)
\int\! \frac{d^4 \ell}{(2\pi)^4}\
\frac{1}{[1 +  \ell^{\mu} P_{\mu\nu}\ell^{\nu}]^{N+1}} 
\\
&
\times
\int_0^1\! dx^1\cdots \int _0^1\! dx^N\
\theta\!\left(\sum_{i=1}^N x^i < 1\right)\,
\left(\sum_{i=1}^N x^i\right)^{N-3}\
\frac{N(l(x,\ell))}
{[\Lambda^{\!2}(x) + \mi 0]^{N-1}}
\;\;.
\label{eq:lxspacemod}
\end{split}
\end{equation}
The function $l(x,\ell)$ in the numerator function is obtained by combining Eqs.~(\ref{eq:tildelinverse}) and (\ref{eq:elldef}): 
\begin{equation}
l^\mu(x,\ell) = 
\frac{1}{1-x^0}
\left[
{\textstyle{\frac{1}{2}}}
\Lambda(x)
\left\{
(1 - \mi)\ell^{ \mu}
+  
(1 + \mi)P^\mu_\nu \ell^{\nu}
\right\}
+\sum_{j=1}^N x^j Q^\mu_j
\right]
\;\;.
\label{eq:elltol}
\end{equation}
It is a somewhat subtle matter to verify that the complex rotations involved in defining $\ell$ are consistent with the $+\mi0$ prescription in the original denominator. We examine this issue in Appendix~\ref{app:wick}.

Notice that in the numerator function the momentum on line $n$ is
\begin{equation}
l^\mu(x,\ell) - Q_n^\mu
=
\frac{1}{1-x^0}
\left[
{\textstyle{\frac{1}{2}}}
\Lambda(x)
\left\{
(1 - \mi)\ell^{ \mu}
+  
(1 + \mi)P^\mu_\nu \ell^{\nu}
\right\}
+K_n^\mu(x)
\right]
\;\;,
\label{eq:numeratorn}
\end{equation}
where
\begin{equation}
K_n^\mu(x) = \sum_{j=1}^N x^j (Q^\mu_j - Q^\mu_n)
\;\;.
\label{eq:Kndef}
\end{equation}
We shall meet $K_n(x)$ later in Sec.~\ref{sec:pinch} when we study pinch singularities. For the moment, we note simply that in the final formula (\ref{eq:lxspacemod}) both the numerator and the denominator are invariant under shifts $Q_i \to Q_i + \Delta Q$ of the reference momenta $Q_i$.

\section{Contour deformation in Feynman parameter space}
\label{deformation}

The integral in Eq.~(\ref{eq:lxspacemod}) is not yet directly suitable for Monte Carlo integration. The problem is that the quadratic function $\Lambda^2(x)$ vanishes on a surface in the space of the Feynman parameters. Evidently, the integrand is singular on this surface. For $x^0 > 0$, $\Lambda^2(x)$ does not vanish for real $x^j$, but on the plane $x^0 = 0$, $\Lambda^2(x)$ vanishes for certain real values of the other $x^j$. If we don't do something about this singularity, the numerical integral will diverge. The something that we should do is deform the integration contour in the direction indicated by the $+\mi 0$ prescription. That is, we write the integral as
\begin{equation}
\begin{split}
{\cal M} ={}& 
- m_0^2 e^N \Gamma(N+1)
\int\! \frac{d^4 \ell}{(2\pi)^4}\
\frac{1}{[1 +  \ell^{\mu} P_{\mu\nu}\ell^{\nu}]^{N+1}}
\int_{C}d z\, 
\left(\sum_{i=1}^N z^i\right)^{\!\!N-3}\
	\frac{N(l(z,\ell))}{\big[\Lambda^{2}(z)\big]^{N-1}}
\\
\equiv{}& 
- m_0^2  e^N \Gamma(N+1)
\int\! \frac{d^4 \ell}{(2\pi)^4}
\frac{1}{[1 +  \ell^{\mu} P_{\mu\nu}\ell^{\nu}]^{N+1}}
\\ & \times
\int_{0}^{1}\!d\xi^{1}\cdots\int_{0}^{1}\!d\xi^{N}\
    \theta\left(\sum_{i = 1}^N \xi^i < 1\right)\,
	\det\!\left(\frac{dz}{d\xi}\right)\,
	\left(\sum_{i=1}^N z^i\right)^{\!\!N-3}\
	\frac{N(l(z(\xi),\ell))}
	{\big[\Lambda^{2}(z(\xi))\big]^{N-1}}
\;\;.
\label{eq:lxspacedeformed}
\end{split}
\end{equation}
Here we integrate over real parameters $\xi^i$ for $i \in \{0,1,\dots,N\}$ with 
$\sum_{i=0}^N \xi^i = 1$, so that we have displayed the integral as an integral over $N$ parameters $\xi^1,\dots,\xi^N$ with $\xi^0 \equiv 1 - \sum_{i = 1}^N \xi^i$. The integration range is $0 < \xi^i$ for $i \in \{0,1,\dots,N\}$. The original integral was over real parameters $x^i$ with $\sum_{i=0}^N x^i = 1$ with this same range, $0 < x^i$. The contour is defined by specifying complex functions $z^i(\xi)$ for $i \in \{0,1,\dots,N\}$ with $\sum_{i=0}^N z^i = 1$.

In moving the integration contour we make use of the multidimensional version of the widely used one dimensional contour integration formula. A simple proof is given in Ref.~\cite{beowulfPRD}. The essence of the theorem is that we can move the integration contour as long as we start in the direction indicated by the $+\mi 0$ prescription and do not encounter any singularities of the integrand along the way. In addition, the boundary surfaces of the contour have to remain fixed. Since the surfaces $z^i = 0$ are boundary surfaces of the contour before deformation, they should remain boundary surfaces after the deformation. The original integral covers the region $0 < x^i$ for $i \in \{1,\dots,N\}$ and the $\xi^i$ cover this same range, $0 < \xi^i$. Thus we demand
\begin{equation}
\label{eq:endpoints1}
z^i(\xi) \to 0 \hskip 1 cm {\rm as}\ \xi^i \to 0
\end{equation}
for $i \in \{0,1,\dots, N\}$.

We adopt a simple ansatz for the contour in the complex $z$-space:\footnote{This is a non-trivial deformation for all $\xi$ such that $\eta(\xi) \ne 0$ with one exception. If all of the $\xi^i$ vanish except for $\xi^n$, where then $\xi^n = 1$, then $z^i = \xi^i$ for all $i$ even if $\eta^n \ne 0$. This possibility does not cause any problems.}
\begin{equation}
\label{eq:contour}
z^{i}(\xi) = \frac{\xi^{i}+\mi \eta^{i}(\xi)}
{1+\mi\sum_{j=0}^N\eta^{j}(\xi)}
\;\;.
\end{equation}
Here the $\eta^{i}$ variables are functions of the integration parameters $\xi^{i}$. With this ansatz, the constraint that $\sum_i z^i = 1$ is automatically satisfied:
\begin{equation}
\sum_{i=0}^N \xi^{i} = 1\qquad \Longrightarrow\qquad 
\sum_{i=0}^N z^{i} = 1
\;\;.
\end{equation}
In order to satisfy Eq.~(\ref{eq:endpoints1}), we require
\begin{equation}
\label{eq:endpoints2}
\eta^i(\xi) \to 0 \hskip 1 cm {\rm as}\ \xi^i \to 0
\end{equation}
for $i \in \{0,1,\dots, N\}$.

There are certain conditions to be imposed on the contour choice in order
to be consistent with the ``$+\mi 0$'' prescription in the original integral. Note first that
\begin{equation}
\label{eq:LambdaFactors}
\Lambda^2(z) = \frac{\Lambda^2(\xi + \mi  \eta(\xi))}
{(1 + \mi \sum_{j=0}^N \eta^j(\xi))^2}
\;\;.
\end{equation}
Next, note that $\Lambda^2$ with argument $\xi + \mi \eta$ appears in the
numerator. In order to analyze Eq.~(\ref{eq:LambdaFactors}), it is convenient to give a special name ${\cal S}(x)$ to the quadratic function that forms the first part of $\Lambda^2$ in Eq.~(\ref{eq:Lambdasqdef}),
\begin{equation}
\Lambda^2(x) = {\cal S}(x) + \mi\sum_{j=1}^N x^0  x^j \ m_0^2
\;\;,
\end{equation}
where
\begin{equation}
{\cal S}(x) = \frac{1}{2} \sum_{i,j = 1}^N x^i x^j S_{ij}
\;\;.
\end{equation}

A sufficient condition for the choice of the $\eta^i(\xi)$ is as follows. 
First, we choose
\begin{equation}
\label{eq:eta0}
\eta^0 = 0
\;\;.
\end{equation}
This is the simplest way to satisfy Eq.~(\ref{eq:endpoints2}) for $\eta^0$. With this choice for $\eta^0$, we have
\begin{equation}
\label{eq:Lambdaexpansion}
\Lambda^2(\xi + \mi \eta) = 
{\cal S}(\xi) - {\cal S}(\eta) 
- m_0^2\,\xi^0 \sum_{j=1}^N \eta^j
+ \mi \sum_{i=1}^N  \eta^i(\xi)\, w_{i}(\xi)
+ \mi m_0^2\,
\xi^0(1-\xi^0)
\;\;,
\end{equation}
where
\begin{equation}
w_i(\xi) \equiv  \frac{\partial {\cal S}(\xi)}{\partial \xi^i}
= \sum_{j=1}^N S_{ij}\xi^j
\;\;.
\end{equation}
Our condition for the choice of the $\eta^i$ for $i \in \{1,\dots,N\}$ is that
\begin{equation}
\label{eq:posimag}
\sum_{i=1}^N \eta^i(\xi) w_i(\xi) \ge 0
\;\;,
\end{equation}
with $\sum \eta^i w_i > 0$ except at a point on the boundary of the integration region. 

Suppose, now, that the condition (\ref{eq:posimag}) is satisfied. Do we then have an allowed contour deformation? Consider the family of contour deformations $\eta^i(\xi;\lambda) = \lambda\, \eta^i(\xi)$ with $0 < \lambda \le 1$. 

We first consider infinitesimal values of $\lambda$. We have, to first
order in $\lambda$,
\begin{equation}
\begin{split}
\Lambda^2(z) = {}& \left[
{\cal S}(\xi) 
- \lambda m_0^2\,\xi^0 \sum_{j=1}^N \eta^j
+ \mi \lambda \sum_{i=1}^N \eta^i(\xi)\, w_{i}(\xi)
+\mi m_0^2\, \xi^0(1-\xi^0)\right]
\\ & \times
\left[
1 - 2\mi \lambda \sum_{j=1}^N \eta^j(\xi)
\right] 
+ {\cal O}(\lambda^2)
\\
= {} &
{\cal S}(\xi)
- \lambda m_0^2\,\xi^0 \sum_{j=1}^N \eta^j
+ 2 \lambda \xi^0 (1 - \xi^0) m_0^2 \sum_{j=1}^N \eta^j
\\&
+ \mi \lambda \sum_{i=1}^N \eta^i(\xi)\, w_{i}(\xi)
- 2\mi \lambda {\cal S}(\xi) \sum_{j=1}^N \eta^j(\xi)
+ \mi m_0^2\, \xi^0(1-\xi^0)
+ {\cal O}(\lambda^2)
\;\;.
\end{split}
\end{equation}
In the neighborhood of any point $\xi$ with $\xi^0 >0$, $\Lambda^2(z)$ has a positive imaginary part even with $\lambda = 0$. For $\xi^0 = 0$, the contour deformation gives $\Lambda^2(z)$ a positive imaginary part in a neighborhood of any point $\xi$ where the real part, ${\cal S}(\xi)$, vanishes. This is the meaning of the ``$+ \mi 0$'' prescription. We may consider that we start with a value of $\lambda$ that is just infinitesimally greater than zero, so that the contour does not actually pass through any poles of the integrand in the interior of the integration region.

Now we turn to larger values of $\lambda$. We have
\begin{equation}
\Lambda^2(z) = \frac{\Lambda^2(\xi + \mi \lambda \eta(\xi))}
{(1 + \mi \lambda \sum_j \eta^j(\xi))^2}
\;\;.
\end{equation}
Assuming that the $\eta^i(\xi)$ are smooth functions, this is a smooth
function of $\xi$. (Note here that $1 + \mi \lambda \sum_j \eta^j(\xi)$
cannot vanish because its real part is 1). Furthermore, when $\lambda > 0$, $\Lambda^2(z)$ is never zero in the interior of the integration region. This is because, according to Eq.~(\ref{eq:Lambdaexpansion}), the imaginary part of $\Lambda^2(\xi + \mi \lambda \eta(\xi))$ is positive. Thus $1/[\Lambda^2(z)]^{N-1}$ is an analytic function of $z$ in the interior of the entire region covered by the family of deformations. For the boundary of the integration region, there are some issues of convergence that one should check. We do so in Appendix \ref{app:DeformCheck}. Anticipating the result of this check, we conclude that the integral is independent of the amount of deformation and we can set $\lambda = 1$.

It is remarkable that the imaginary part of $\Lambda^2(z)$ is not
necessarily positive on all of the deformed contour. What is crucial is
that the deformation starts in the right direction and that, as the
contour is deformed, it does not cross any poles.

\section{A standard contour deformation}
\label{standarddeformation}

A convenient choice for the deformation function $\eta^i(\xi)$ for $i \in \{1,\dots,N\}$ is
\begin{equation}
\label{eq:etadef}
\eta^i(\xi) = (\lambda/m^2)\, \xi^i w_i(\xi)
\;\;.
\end{equation}
Here $\lambda$ is an adjustable dimensionless constant and $m^2$ is a
parameter with the dimension of squared mass that we insert because
$S_{ij}$ and thus $w_i$ has dimension of squared mass. Note that with this choice the requirement~(\ref{eq:endpoints2}) that $\eta^i(\xi)$ vanish when $\xi^i$ vanishes is automatically met. This deformation gives
\begin{equation}
{\cal S}(\xi + \mi\eta) = 
{\cal S}(\xi) - {\cal S}(\eta) 
+ \mi (\lambda/m^2)\sum_{i=1}^N \xi^i\, [w_i(\xi)]^2
\;\;.
\end{equation}
Evidently the imaginary part of $\Lambda^2(\xi + \mi\eta)$ has the right
sign.

Eq.~(\ref{eq:etadef}) can be thought of as specifying a basic deformation. We can add other deformations to this. In our numerical work for this paper we have added one more deformation, as specified in Appendix~\ref{app:extradeform}.

\section{Pinch singularities}
\label{sec:pinch}

The integrand is singular for $\xi^0 \to  0$ at any real point $\xi$ with ${\cal S}(\xi) = 0$. We have seen in the previous section that the standard contour deformation keeps the contour away from this singularity as long as there is some index $i \in \{1,\dots,N\}$ such that $\xi^i > 0$ and $w_i(\xi) \ne 0$.

What about a point $\xi$ with ${\cal S}(\xi) = 0$ such that there is {\it no} index $i \in \{1,\dots,N\}$ such that $\xi^i > 0$ and $w_i(\xi) \ne 0$. In this case, the integration contour is pinched in the sense that there is no allowed contour deformation that can give ${\cal S}(\xi+ \mi \eta)$ a positive imaginary part at this point $\xi$. To see this, recall from Eq.~(\ref{eq:Lambdaexpansion}) that, when $\xi^0 = 0$,
\begin{equation}
\label{eq:ImLambda}
{\rm Im}\,{\cal S}(\xi + \mi \eta) = \sum_{i=1}^N \eta^i w_i(\xi)
\;\;.
\end{equation}
Consider a point $\xi$ such that ${\cal S}(\xi) = 0 $ and such that for each index $i \in \{1,\dots,N\}$, $w_i(\xi) \ne 0$ implies $\xi^i = 0$. For any allowed deformation $\eta(\xi)$, we must have $\eta^i = 0$ for all $i \in \{1,\dots,N\}$ such that $\xi^i = 0$. Thus for each index $i \in \{1,\dots,N\}$, $w_i(\xi) \ne 0$ implies $\eta^i = 0$. From  Eq.~(\ref{eq:ImLambda}) we conclude that ${\rm Im}\,{\cal S}(\xi + \mi \eta)$ must vanish at the point in question for any allowed choice of the $\eta^i$. We conclude that a real point $\xi$ with $\xi^0 = 0$ and with ${\cal S}(\xi) = 0$ is a pinch singular point if, and only if, 
\begin{equation}
\label{eq:pinchcondition1}
\xi^i w_i(\xi) = 0\hskip 1 cm {\rm for\ every}\ i \in \{1,\dots,N\}
\;\;.
\end{equation}

We also note that the point $\xi^0 = 1$, $\xi^i = 0$ for $i \in \{1,\dots,N\}$, is a pinch singular point. This singularity corresponds to the ultraviolet region of the original loop integration. 

With a little algebra, one can translate the condition for a pinch singularity with $\xi^0 = 0$. At one of these points, we have
\begin{equation}
\label{eq:pinchcondition2}
{\rm either}\quad \xi^i = 0\quad{\rm or}\quad K_i^2 - m_i^2 = 0
\end{equation}
for each $i \in \{1,\dots,N\}$, where $K^\mu_i(\xi)$ was given earlier in Eq.~(\ref{eq:Kndef}). When $\xi^0 = 0$, these vectors have the properties that $K_i - K_{i+1} = P_i$ and $\sum \xi_i K_i = 0$. Thus Eq.~(\ref{eq:pinchcondition2}) is the well known condition for a pinch singularity (see Bjorken and Drell \cite{bjorkenanddrell}). It says that for each propagator $i$ around the loop, there is a momentum $K_i$ such that momentum conservation is obeyed at the vertices and each $K_i$ around the loop is either on shell or else the corresponding $\xi^i$ is zero and such that the space-time separations $\Delta x_i^\mu = \xi^i K_i^\mu$ around the loop sum to zero.

Notice that the momenta $K^\mu_i(\xi)$ appear in the numerator function. According to Eq.~(\ref{eq:numeratorn}), the momentum for line $i$ in the numerator function in the case that $\xi$ is at a contour pinch (so $\Lambda(\xi) = 0$) is $K^\mu_i(\xi)$.

There are two types of pinch singular points that are always present if we have massless kinematics (with no external momenta collinear to each other) and one more that can be present. 

\subsection{Soft singularity}
\label{sec:soft}

The first kind of pinch singular point that is always present if we have massless kinematics is the one corresponding to a loop propagator momentum that vanishes. If $m_n = 0$ for some $n$ then $S_{nn} = 0$. This means that $\Lambda^2(\xi) = 0$ when all of the $\xi^i$ vanish except for $\xi^n$, which is then $\xi^n = 1$. This is a pinch singular point because all of the $z^i(\xi)$ are fixed: $z^i = 0$ for $i \ne n$, $z^n = 1$. This point corresponds to the momentum of line $n$ in the momentum space representation vanishing. In our photon scattering example, there is a singularity at this point but because of the zero from the numerator function it is not strong enough to produce a divergence.

\subsection{Collinear singularity}
\label{sec:collinear}

The second kind of pinch singular point that is always present if we have massless kinematics is the one corresponding to two loop propagator momenta becoming collinear to an external momentum. If $m_n = m_{n+1} = 0$ for some $n$ and if $P_n^2 = (Q_{n+1} - Q_n)^2 = 0$, then $S_{nn} = S_{n+1,n+1} = S_{n+1,n} = 0$. This means that $\Lambda^2(\xi) = 0$ when all of the $\xi^i$ vanish except for $\xi^n$ and $\xi^{n+1}$. It also means that $w_n(\xi) = w_{n+1}(\xi) = 0$, so that this is a pinch singular point according to the condition (\ref{eq:pinchcondition1}). This point corresponds to the momentum of lines $n$ and $n+1$ in the momentum space representation being collinear with $P_n$. In our photon scattering example, there is a singularity along this line but because of the zero from the numerator function it is not strong enough to produce a divergence.

\subsection{Double parton scattering singularity}
\label{sec:dps}

\begin{figure}
\centerline{
\includegraphics[width = 10 cm]{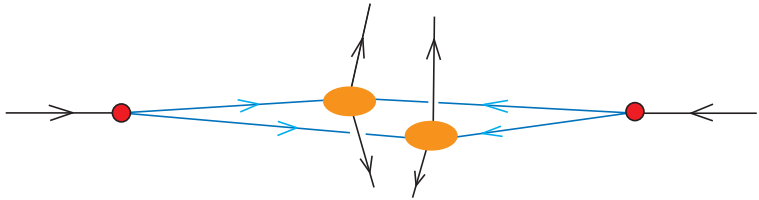}
}
\medskip
\caption{
Illustration of the double parton scattering singularity.
}
\label{fig:dps}
\end{figure}

A third type of pinch singular point can be present if a special condition holds for the external momenta. This singularity corresponds to double parton scattering and is illustrated in Fig.~\ref{fig:dps}. Imagine that incoming parton $A$ splits into two collinear partons. Imagine also that incoming parton with index $B$ splits into two collinear partons. One of the partons from $A$ and one from $B$ could meet and produce a group of final state partons. The other parton from $A$ could meet the other parton from $B$ and produce a second group of final state partons. For this to happen, we need at least two external lines in each group of outgoing partons produced. Thus we need at least four outgoing external particles. Thus we need $N \ge 6$.

This picture satisfies the criteria of Eq.~(\ref{eq:pinchcondition2}) for a pinch singularity. In the Feynman parameter space, the singularity occurs along a one dimensional line in the interior of the space. We work out where this line is in Appendix \ref{sec:appendixdps}.
 
Now, the pinch singularity conditions hold only for certain special choices of the external momenta. However, if $N$ is large, it is usual that the kinematics is close to a pinch singularity condition for some of the graphs. For this reason, in a numerical program, one should check for each graph if such a nearly pinched contour occurs. In the event that it does, one should put a high density of integration points near the almost singular line.

\section{Ultraviolet subtraction}
\label{sec:UV}

Some graphs are ultraviolet divergent. For instance, in the photon scattering case, there is an ultraviolet divergence for $N = 4$ (and for $N = 2$, but we do not consider that case.) In the representation (\ref{eq:lxspacedeformed}), the divergence appears as a divergence from the integration over Feynman parameters near $\xi^0 = 1$, with all of the other $\xi^i$ near zero. The reader can check that with a numerator function proportional to $N$ powers of the loop momentum, this region does give a logarithmic divergence for $N = 4$. If the graph considered is ultraviolet divergent, it needs an ultraviolet subtraction, so that we calculate
\begin{equation}
{\cal M}_{\rm net} = {\cal M} - {\cal M}_{\rm uv}
\;\;.
\end{equation}
In a numerical integration, we subtract the integrand of ${\cal M}_{\rm uv}$ from the integrand of ${\cal M}$, then integrate. We arrange that the singularities of the integrand cancel to a degree sufficient to remove the divergence. The subtraction term is defined in Ref.~\cite{NSsubtractions} so that it reproduces the result of $\overline{\rm MS}$ subtraction (which we use because it is gauge invariant). In the photon scattering case, the sum over graphs of the subtraction term vanishes, corresponding to the fact that there is no elementary four photon vertex. Thus the result after summing over graphs does not depend on the $\overline{\rm MS}$ renormalization scale.

The subtraction term from Eq.~(A.37) of Ref.~\cite{NSsubtractions} is
\begin{equation}
\begin{split}
{\cal M}_{\rm uv} ={}& 
\int\! \frac{d^4 l}{(2\pi)^4}\
\frac{\mi m_0^2\,e^4 }{\mi m_0^2}
\Bigg\{
\frac{N(l,l,l,l)- 32 \prod_{j=1}^4 l\cdot \epsilon_j(P_j)}
{ [l^2 - \mu^2 e^{-4/3}+ \mi 0]^4}
+\frac{32 \prod_{j=1}^4 l\cdot \epsilon_j(P_j)}
{ [l^2 - \mu^2 e^{-3/2}+ \mi 0]^4}
\Bigg\}
\;\;.
\label{eq:UVsubtraction}
\end{split}
\end{equation}
Here $\mu^2$ is the $\overline{\rm MS}$ renormalization scale, which can be anything we like since the net counter-term is zero. For the numerator function in the first term, we have adopted the notation that the ordinary numerator function $N(l)$ in Eq.~(\ref{eq:numeratordef}) is written $N(k_4,k_3,k_2,k_1)$ where $k_n = l - Q_n$. In this notation, $N(l,l,l,l)$ is the standard numerator function with each propagator momentum set equal to $l$.

We can now apply the same transformations as for the starting graph to obtain the representation
\begin{equation}
\begin{split}
{\cal M}_{\rm uv} ={}&
- m_0^2 e^4 \Gamma(5)
\int\! \frac{d^4 \ell}{(2\pi)^4}\
\frac{1}{[1 +  \ell^{\mu} P_{\mu\nu}\ell^{\nu}]^{5}} 
\\
& \times
\int_0^1\! dx^1\cdots \int _0^1\! dx^4\
\theta\!\left(\sum_{i=1}^4 x^i < 1\right)\,
\left(\sum_{i=1}^4 x^i\right)
\\
& \times
\Bigg\{
\frac{N(l,l,l,l)- 32 \prod_{j=1}^4 l\cdot \epsilon_j(P_j)}
{[\Lambda_{4/3}^{\!2}(x) + \mi 0]^{3}}
+
\frac{32 \prod_{j=1}^4 \tilde l\cdot \epsilon_j(P_j)}
{[\Lambda_{3/2}^{\!2}(x) + \mi 0]^{3}}
\Bigg\}
\;\;.
\label{eq:lxspaceUV}
\end{split}
\end{equation}
Here
\begin{equation}
\begin{split}
\Lambda_{4/3}^{\!2}(x) ={}&
- (1-x^0)^2 \mu^2 e^{-4/3} + \mi x^0(1-x^0) m_0^2
\;\;,
\\
\Lambda_{3/2}^{\!2}(x) ={}&
- (1-x^0)^2 \mu^2 e^{-3/2} + \mi x^0(1-x^0) m_0^2
\;\;,
\end{split}
\end{equation}
where we have used $x^0 = 1 - \sum_{j=1}^4 x^j$. In the numerator of the first term, $l$ is a function $l(x,\ell)$,
\begin{equation}
l^\mu(x,\ell) = \frac{1}{1-x^0}
\left[
{\textstyle{\frac{1}{2}}}
\Lambda_{4/3}(x)
\left\{
(1 - \mi)\ell^{ \mu}
+  
(1 + \mi)P^\mu_\nu \ell^{\nu}
\right\}
\right]
\;\;.
\label{eq:elltolUV1}
\end{equation}
In the second term, $\tilde l$ in the numerator is a function $\tilde l(x,\ell)$,
\begin{equation}
\tilde l^\mu(x,\ell) = \frac{1}{1-x^0}
\left[
{\textstyle{\frac{1}{2}}}
\Lambda_{3/2}(x)
\left\{
(1 - \mi)\ell^{ \mu}
+  
(1 + \mi)P^\mu_\nu \ell^{\nu}
\right\}
\right]
\;\;.
\label{eq:elltolUV2}
\end{equation}

The reader can check that for the photon scattering case with $N=4$ the integrand for ultraviolet subtraction matches that of the starting graph in the region $x^0 \to 1$, so that if we subtract the integrand from the counter-term graph from the integrand for the starting graph, the resulting integral will be convergent.

\section{The Monte Carlo Integration}
\label{sec:MonteCarlo}

We have implemented the integration in Eq.~(\ref{eq:lxspacedeformed}) as computer code \cite{whereiscode}. The integration is performed by the Monte Carlo method. This is a standard method, but it may be good to indicate what is involved. First, we note that we do not simply feed the integrand to a program that can integrate ``any'' function. There are many reasons for this, but the most important is that we do not have just {\em any} function but a function with a known singularity structure, a structure that is generic to loop diagrams in quantum field theory with massless kinematics. We can take advantage of our knowledge of how the integrand behaves. 

To proceed, we note that we have an integral of the form
\begin{equation}
{\cal M} = \int\! d^4 \ell\
\int_0^1 \!d\xi^0 \int_0^1\!d\xi^1 \cdots \int_0^1\!d\xi^N
\delta\!\left(\sum_{i=0}^{N} \xi^i - 1\right)\
f(\ell,\xi)\;\;.
\label{integralform}
\end{equation}
In a Monte Carlo integration, we choose $N_{\rm pts}$ points $\{\ell_j,\xi_j\}$ at random with a density $\rho(\ell,\xi)$ and evaluate the integrand $f(\xi)$ at these points. Then the integral is
\begin{equation}
{\cal M} = \lim_{N_{\rm pts} \to \infty} \frac{1}{N_{\rm pts}}
\sum_{j=1}^{N_{\rm pts}}
\frac{f(\ell_j,\xi_j)}{\rho(\ell_j,\xi_j)}
\;\;.
\end{equation}
The integration error with a finite number of points is proportional to $1/\sqrt{N_{\rm pts}}$. The coefficient of $1/\sqrt{N_{\rm points}}$ in the error is smallest if 
\begin{equation}
\rho(\ell,\xi) \approx {\it const.}\times |f(\ell,\xi)|
\;\;.
\end{equation}
That is the ideal, but it is not really possible to achieve this ideal to the degree that one has a one part per mill error with one million points. However, one would certainly like to keep $|f(\ell,\xi)|/\rho(\ell,\xi)$ from being very large. In particular, $f(\ell,\xi)$ is singular along certain lines in the space of the $\xi$ (the collinear singularities) and at certain points (the soft singularities). We need to arrange that $\rho$ is singular at the same places that $f$ is singular, so that ${f(\ell,\xi)}/{\rho(\ell,\xi)}$ is {\em not} singular anywhere. Since $|f(\ell,\xi)|$ can be very large near other lines associated with double parton scattering, we also need to arrange that $\rho(\ell,\xi)$ is similarly large near these lines.

We construct the desired density in the form
\begin{equation}
\rho(\ell,\xi) = \rho_\ell(\ell)
\sum_{J=1}^{N_{\rm alg}} \alpha_J\,\rho_J(\xi)
\;\;.
\end{equation}
Here $\int\! d^4\ell\, \rho_\ell(\ell) = 1$, the sum of the $\alpha_J$ is 1, and there are several density functions $\rho_J$ with
$\int\! d\xi\, \rho_J(\xi) = 1$. Each $\rho_J$ corresponds to a certain algorithm for choosing a point $\xi$. For each new integration point, the computer chooses which algorithm to use with probability $\alpha_J$.
The various sampling algorithms are designed to put points into regions in which the denominator is small, based on the coefficients $S_{ij}$. We omit describing the details of the sampling methods since these are likely to change in future implementations of this style of calculation.

Points $\ell$ are chosen with a simple distribution $\rho_\ell(\ell)$. In the calculation of the numerator, we average between the numerator calculated with $\ell$ and the numerator calculated with $-\ell$. 

Having outlined how $\cal M$ is calculated by Monte Carlo integration, we pause to suggest how the calculation of a cross section (for, say, Higgs production) would work. There one would have one-loop amplitudes $\cal M$ expressed as integrals and one would need to multiply $\cal M$ by a function $h(P)$ of the external momenta that represents a tree amplitude and a definition of the observable to be measured. One would need the integral of this over the external momenta $P$. One would perform all of the integrations together. That is, one would choose points $\{P,\ell,\xi\}$ and calculate the contributions from the virtual graphs times tree graphs to the desired cross section according to 
\begin{equation}
I = \lim_{N_{\rm pts} \to \infty} \frac{1}{N_{\rm pts}}
\sum_{j=1}^{N_{\rm pts}}
\frac{f(\ell_j,\xi_j;P_j)}{\rho(\ell_j,\xi_j,P_j)}\ h(P_j)
\;\;.
\end{equation}
Here $f$ is the integrand of $\cal M$ as above. The function $\rho$ is the net density of points in $\ell$, $\xi$, and $P$. Thus what one would use is not $\cal M$ itself but rather the integrand for $\cal M$.

\section{Checks on the calculation}
\label{sec:checks}

As discussed in the previous section, we have implemented the integration in Eq.~(\ref{eq:lxspacedeformed}) as computer code \cite{whereiscode}. With this code, there are a number of internal checks that can be performed on the computation. First, we can replace the real integrand by a function that has soft or collinear singularities but is simple enough to easily integrate analytically. Then we can compare the numerical result to the analytical result. This checks that the functions $\rho_i(\xi)$ and $\rho(\ell)$ correspond to the true probabilities with which points $\xi$ and $\ell$ are chosen. Then we can vary the amount of deformation (both for the real integral and for the test integrals). When we integrate over a different contour, the integrand is quite different. Nevertheless, the $(N+4)$-dimensional Cauchy theorem guarantees that the integral should be unchanged, provided that the integration is being performed correctly. Thus invariance under change of contour is a powerful check. Another check is to change the value of the parameter $m_0$. At the start, $\cal M$ is proportional to $m_0^2/m_0^2$ and is trivially independent of $m_0$. However in the integral as performed, Eq.~(\ref{eq:lxspacedeformed}), $m_0$ is deeply embedded into the structure of the integrand, so that it is a non-trivial check on the integration that the result does not change when we change $m_0$. Next, we can replace one of the photon polarization vectors $\epsilon_n(P_n)$ by $P_n$. This gives a non-zero result for each Feynman graph, but should give zero for the complete amplitude summed over graphs. Additionally, we can change the the definition of the polarization vectors $\epsilon_n(P_n)$. For reasons of good numerical convergence, we normally use polarization vectors appropriate for Coulomb gauge, but we can switch to a null-plane gauge. The two amplitudes should differ by a phase, so that $|{\cal M}|$ is unchanged. Another check is obtained by replacing the vector current by a left-handed current or a right-handed current. For even $N$, the left-handed and right-handed results should be the same, while for odd $N$ they should be opposite. For another check, we can reformulate the integral so that we do not define $\ell$ with a scale $\Lambda(x)^{-1}$. In this formulation, the denominator is 
\begin{equation}
[{\cal S}(x) 
+ \mi x^0(1-x^0) m_0^2 (1 + \ell^{\mu} P_{\mu\nu}\ell^{\nu})]^{N+1}
\;\;.
\end{equation}
The structure of the integral is quite different, but the result should be the same. For four photons, there is one additional test: the result should be independent of the renormalization parameter $\mu$.

We have subjected the code \cite{whereiscode} to these checks. The numerical precision of the result is often not high and we have not used every check for every choice of $N$ and external momenta and polarizations. Nevertheless, we have found that where we have tried them, the various checks are always passed. We note that a better check would be to obtain the same results with completely independently written code. We have not done that. 

\section{Results}
\label{sec:results}

In this section, we use this code to test how well the method described works. The result for a given choice of helicities of the photons has a phase that depends on the precise definition of the photon polarization vectors $\epsilon_i$. However, the absolute value of the scattering amplitude ${\cal M}$ is independent of this conventions used to define the $\epsilon_i$, so we concentrate on $|{\cal M}|$. Our convention for defining ${\cal M}$ is specified in Eq.~(\ref{eq:lspace0}). Since $|{\cal M}|$ is proportional to $\alpha^{N/2}$ and has mass dimension $4-N$, we exhibit $|{\cal M}|\times (\sqrt{s})^{N-4}/\alpha^{N/2}$ in our plots. We specify helicities in the form $h_1,h_2,h_3,\dots,h_N$, where 1 and 2 are the incoming particles and, following convention, $h_1$ and $h_2$ are actually the negative of the physical helicities of the incoming photons.

\subsection{$N = 4$}
\label{sec:results4}
\begin{figure}
\centerline{
\includegraphics[width = 10 cm]{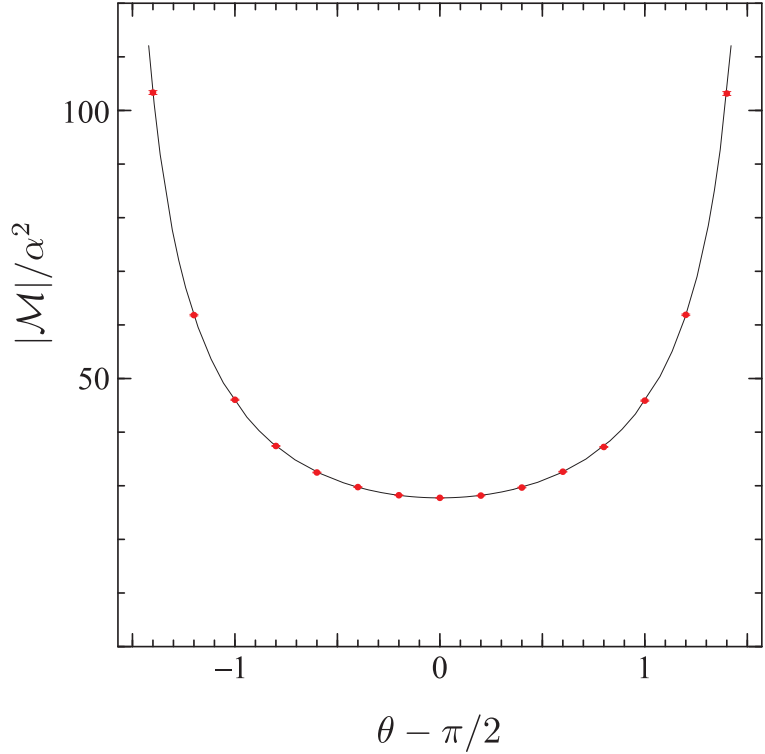}
}
\caption{
Four photon amplitude. We plot $|{\cal M}|/\alpha^{2}$ for helicities $+$$+$$-$$-$ versus $\theta - \pi/2$, where $\theta$ is the scattering angle. The curve is the analytical result from Ref.~\cite{lightbylight}. The points are the result of numerical integration.
}
\label{fig:Fourphotons}
\end{figure}

We begin with $N = 4$, light-by-light scattering. Here we use the subtraction for the ultraviolet divergence in each graph as described in Sec.~\ref{sec:UV}. For the $N=4$ case the result is known and has been presented in a convenient form in Ref.~\cite{lightbylight}. For the two helicity choices $+$$+$$+$$+$ and $+$$+$$+$$-$, $|{\cal M}|/\alpha^2 = 8$. Our numerical results agree with this. For the choice $+$$+$$-$$-$, the result depends on the value of the scattering angle $\theta$. In Fig.~\ref{fig:Fourphotons} we exhibit the prediction of Ref.~\cite{lightbylight} versus $\theta - \pi/2$ as a curve and a selection of points obtained by numerical integration as points with error bars. The error bars represent the statistical uncertainty in the Monte Carlo integration. It is not easy to see the error bars in the figure. The fractional errors range from 0.0022 to 0.0034. The points were generated using $10^6$ Monte Carlo points for each of six graphs.

\subsection{$N = 5$}
\label{sec:results5}

\begin{figure}
\centerline{
\includegraphics[width = 10 cm]{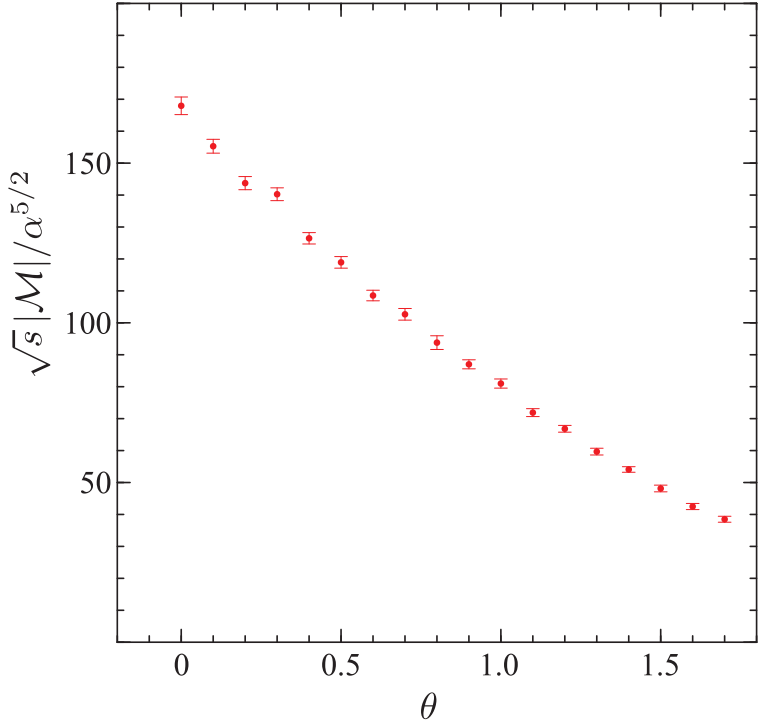}
}
\medskip
\caption{
Five vector boson amplitude. We plot $\sqrt s\,|{\cal M}|/\alpha^{5/2}$ with helicities $+$$-$$-$$+$$+$. The vector boson is massless and couples to the electron with the left-handed part of the photon coupling. An arbitrarily chosen final state was rotated about the $y$-axis through angle $\theta$.
}
\label{fig:Fivephotons}
\end{figure}

We turn next to $N=5$. Since the five photon matrix element vanishes after summing over graphs, we use a massless vector boson that couples to the electron with the left-handed part of the photon coupling. The final state phase space has four dimensions, which does not lend itself to a simple plot. Accordingly we have chosen an arbitrary point for the final state momenta $\{\vec p_3, \vec p_4, \vec p_5\}$:
\begin{equation}
\begin{split}
\vec p_3 ={}& (33.5,15.9,25.0)
\;\;,
\\
\vec p_4 ={}& (-12.5,15.3,0.3)
\;\;,
\\
\vec p_5 ={}& (-21.0,-31.2,-25.3)
\;\;.
\end{split}
\end{equation}
We take photon 1 to have momentum $\vec p_1$ along the $-z$-axis (so the physical incoming momentum is along the $+z$-axis), and we take $\vec p_2$ along the $+z$-axis. Then we create new momentum configurations by rotating the final state through angle $\theta$ about the $y$-axis. In Fig.~\ref{fig:Fivephotons}, we plot computed values of $\sqrt s\,|{\cal M}|/\alpha^{5/2}$ versus $\theta$.
The points were generated using $10^6$ Monte Carlo points for each of 24 graphs.

\subsection{$N = 6$}
\label{sec:results6}

\begin{figure}
\centerline{
\includegraphics[width = 10 cm]{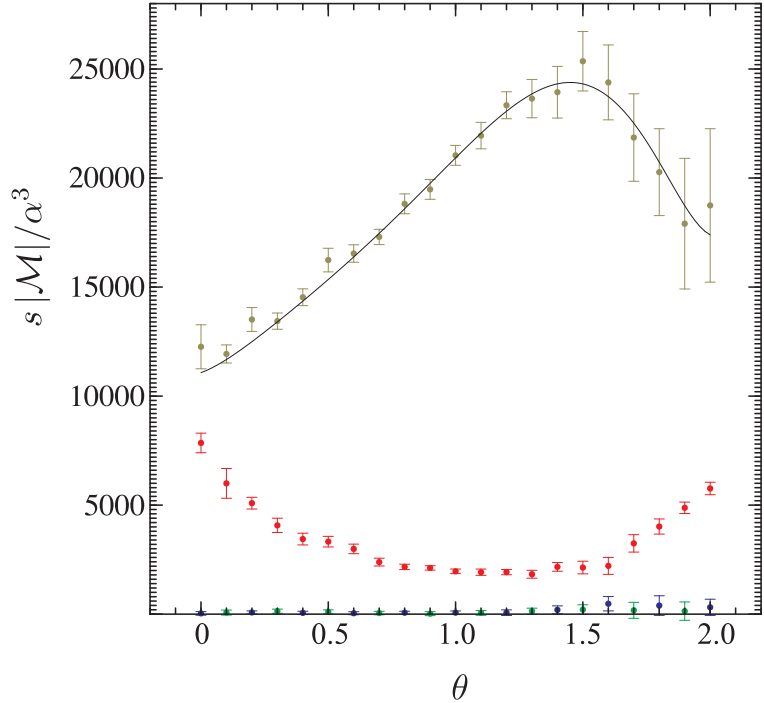}
}
\medskip
\caption{
Six photon amplitude. We plot $s\,|{\cal M}|/\alpha^{3}$. An arbitrarily chosen final state was rotated about the $y$-axis through angle $\theta$. The points are the result of numerical integration. At the  top, the points in the range 10000 to 25000 are for helicities $+$$+$$-$$-$$-$$-$ and are compared with the analytical results of Ref.~\cite{MahlonTahoe}. In the middle, the points in the range from 2000 to 8000 are for helicities $+$$-$$-$$+$$+$$-$. There is no analytical result for this helicity combination. At the bottom, we show numerical results for helicities $+$$+$$+$$+$$+$$+$ and $+$$+$$+$$+$$+$$-$. According to Ref.~\cite{Mahlon}, the amplitude should vanish for these helicity choices. The results for $+$$+$$+$$+$$+$$+$ are computed at $\theta = 0, 0.2, 0.4, \dots$ while the results for $+$$+$$+$$+$$+$$-$ are computed at $\theta = 0.1, 0.3, 0.5, \dots$. 
}
\label{fig:Sixphotons}
\end{figure}

Finally, we compute the six photon amplitude. Here analytic results are known for the helicity choices $+$$+$$+$$+$$+$$+$ and $+$$+$$+$$+$$+$$-$: for these helicity choices, the amplitude should vanish \cite{Mahlon}. There is also a non-zero analytical result for the choice $+$$+$$-$$-$$-$$-$ \cite{MahlonTahoe}.
We compute $s\,|{\cal M}|/\alpha^{3}$ for these helicity choices and also for $+$$-$$-$$+$$+$$-$, for which we know of no analytic result. Following what we did for $N=5$, we choose an arbitrary point for the final state momenta $\{\vec p_3, \vec p_4, \vec p_5, \vec p_6\}$:
\begin{equation}
\begin{split}
\vec p_3 ={}& (33.5,15.9,25.0)
\;\;,
\\
\vec p_4 ={}& (-12.5,15.3,0.3)
\;\;,
\\
\vec p_5 ={}& (-10.0,-18.0,-3.3)
\;\;,
\\
\vec p_6 ={}& (-11.0,-13.2,-22.0)
\;\;.
\end{split}
\end{equation}
We choose $\vec p_1$ and $\vec p_2$ as we did for $N=5$. Then we create new momentum configurations by rotating the final state through angle $\theta$ about the $y$-axis. In Fig.~\ref{fig:Sixphotons}, we plot computed values of $s\,|{\cal M}|/\alpha^{3}$ versus $\theta$. For $+$$+$$+$$+$$+$$+$ helicities, we compute the amplitude at $\theta = 0, 0.2, 0.4, \dots$. The results are consistent with the known result of zero. For $+$$+$$+$$+$$+$$-$ helicities, we compute the amplitude at $\theta = 0.1, 0.3, 0.5, \dots$. The results are again consistent with zero. For $+$$+$$-$$-$$-$$-$ we compare the numerical results to the analytical results of Ref.~\cite{MahlonTahoe} (top curve) and find good agreement. For the helicity choice $+$$-$$-$$+$$+$$-$, the results lie in the range from 2000 to 8000 and exhibit some variation as the final state momenta are varied. We do not have an analytical curve with which to compare. The points were generated using $10^6$ Monte Carlo points for each of 120 graphs.\footnote{This takes a bit under an hour for each point on one chip of our computer, but we note that computer timings are dependent on the computer and the compiler.}

\section{Conclusions}
\label{sec:conclusions}

In the calculation of cross sections for infrared-safe observables in high energy collisions at next-to-leading order, one must treat the real emission of one parton beyond the Born level and one must include virtual loop corrections to the Born graphs. Most calculations follow the method of Ref.~\cite{ERT}, in which the integration over real emission momenta is performed numerically while the integration over virtual loop momenta is performed analytically. One can, however, perform all of the integrations numerically.\footnote{That it is practical to do so has been demonstrated for the case of three jet production in electron-positron annihilation \cite{beowulfPRL,beowulfPRD}. However, the method used there does not extend well to the hadron-hadron case.}

In one approach to the calculation of loop diagrams by numerical integration, one would use a subtraction scheme \cite{NSsubtractions} that removes infrared and collinear divergences from the integrand in a style similar to that used for real emission graphs. Then one would perform the loop integration by Monte Carlo integration along with the integrations over final state momenta. In this paper, we have explored how one would perform the numerical integration. We have studied the $N$-photon scattering amplitude with a massless electron loop in order to have a case with a singular integrand that is not, however, so singular as to require the subtractions of Ref.~\cite{NSsubtractions}.

One could perform the integration either directly as an integral $\int d^4 l$ or with the help a different representation of the integral. We have chosen to explore the use of the Feynman parameter representation because it makes the denominator simple. We have found that this method works for the cases of 4, 5, or 6 external legs. There is, in principle, no limitation to the number of external legs. However, for more external legs, the integrand becomes more singular because the denominator is raised to a high power, $N-1$. This is evident in our results by examining the growth of the integration error as $N$ increases. 

In many practical calculations, the partons in the loop can have non-zero masses and the partons entering the loop can be off-shell. These possibilities make the analytical results more complicated, but we expect that they make the numerical result more stable by softening the singularities.\footnote{Having an unstable massive particle as an incoming parton would be an exception, since this can put new kinds of singularities into the integrand.} However, we leave exploration of this issue for later work.

It is remarkable that the method presented here works for quite a large number of external legs. However, we expect that the method can be improved. One approach lies in making a sequence of small improvements that together amount to a big improvement. Along these lines, one can work on the algorithm for deforming the integration contour and on the sampling methods used for choosing integration points (which methods we have not discussed here). Alternatively, one can look for a different representation of $\cal M$ as an integral. One could use an integral transformation other than that provided by the Feynman parameter representation or one could use a more direct representation of the integral. In particular, the representation of Refs.~\cite{beowulfPRL,beowulfPRD} recommends itself. Here we turn the integral $\int d^4 l$ into a three dimensional integral $\int d^4 \vec l\ \delta_+(l^2)$ that is rather similar to what one has for real parton emissions. In contrast to Refs.~\cite{beowulfPRL,beowulfPRD}, however, one would use explicit subtractions.\footnote{We understand that S.~Catani, T.~Gleisberg, F.~Krauss, G.~Rodrigo, and J.~Winter are working along these lines \cite{Catani}.} We expect that this method or something similar might be superior to the Feynman parameter method used in this paper because for large $N$ one does not raise a denominator to a high power.

The challenge would be to find a representation that is simple and for which the integrand is well enough behaved that one can get numerical results for, say, $N=12$. We hope that others might accept this challenge.

\acknowledgements

This work was supported in part the United States Department of Energy
and by the Swiss National Science Foundation (SNF) through grant no.\
200020-109162 and by the Hungarian Scientific Research Fund grants OTKA
K-60432. We thank T.~Binoth for encouraging us to try the Feynman parameter representation. We thank G.~Heinrich for pointing us toward Ref.~\cite{lightbylight} and L.~Dixon  for pointing us toward Ref.~\cite{MahlonTahoe}.

\appendix 

\section{Wick rotation}
\label{app:wick}

In this appendix, we explain the contour deformation necessary to obtain the scaled and rotated momenta of Eq.~(\ref{eq:elldef}). The simplest procedure is to start by rotating the space-parts of the vector $\tilde l$,
\begin{equation}
\tilde l^0 = k^0\;\;,
\hskip 1 cm
\tilde l^j = e^{-\mi \theta} k^j
\hskip 0.5 cm k = 1,2,3
\;\;.
\end{equation}
Here the components of $k$ are real. We start with $\theta = 0$ and increase $\theta$ until $\theta = \pi/2$. Thus we rotate the $\tilde l$ contour. Throughout the rotation, $\tilde l^2$ has a positive imaginary part. At the end, $\tilde l^2 + \Lambda^2(x)$ becomes $k^\mu P_{\mu\nu} k^\nu + \Lambda^2(x)$, where $k^\mu P_{\mu\nu} k^\nu$ is the euclidean square of $k$.

The next step is to rotate all of the components of $k$ by half of the phase of $\Lambda^2(x)$, so that after the rotation, $k^\mu P_{\mu\nu} k^\nu$ has the same phase as $\Lambda^2(x)$ (which itself has a positive imaginary part).

Finally we rescale the components of $k$ by the absolute value of $\Lambda^2(x)$. Thus
\begin{equation}
k^\mu = \Lambda(x) \ell^\mu\;\;.
\end{equation}
The net transformation is that of Eq.~(\ref{eq:elldef}). At all stages, the imaginary part of the denominator is positive.

\section{Contour deformation}
\label{app:DeformCheck}

In this appendix, we exhibit some details of the argument that the integration over Feynman parameters is left invariant by the contour deformation. We start with
\begin{equation}
I = \int\! d\xi \det A\ F(z(\xi,\lambda))
\;\;.
\end{equation}
Here $z(\xi,\lambda)$ specifies the deformed contour,
\begin{equation}
\int\! d\xi = \int_0^1 d\xi^1 \cdots \int_0^1 d\xi^N\
\theta(\sum_{j=1}^N \xi < 1)
\;\;,
\end{equation}
the matrix $A$ is
\begin{equation}
A^i_j = \frac{\partial z^i}{\partial \xi^j}
\;\;,
\end{equation}
and $F(z)$ is the integrand,
\begin{equation}
F(z) = \left(\sum_{i=1}^N z^i\right)^{\!\!N-3}\
	\frac{N(l(z,\ell))}
	{\big[\Lambda^{2}(z)\big]^{N-1}}
	\;\;.
\end{equation}
In order to take care with what happens at the integration boundary, we define a function $R(\xi)$ that measures how far the point $\xi$ is from the boundary,
\begin{equation}
R(\xi) = \min\!\left(\xi^1,\dots,\xi^N, 1 - \sum_{j=1}^N \xi^j
\right)
\;\;.
\end{equation}
The boundary is at $R(\xi)$ = 0, and in general $0 < R(\xi) < 1/(N+1)$. Then
\begin{equation}
I = \lim_{r \to 0} I(r)
\;\;,
\end{equation}
where
\begin{equation}
I(r) = \int\! d\xi\ \theta(R(\xi) > r)\,\det A\ F(z(\xi,\lambda))
\;\;.
\end{equation}

Now let us make a small change $\delta \lambda$ in the deformation parameter. If we can prove that the corresponding change $\delta I$ of the integral vanishes, then the integral is the same for $\lambda = 1$ as it was for an infinitesimal $\lambda$. We calculate $\delta I(r)$ for non-zero $r$. As shown in Ref.~\cite{beowulfPRD}, $\delta I(r)$ is the integral of a total derivative. Thus we get an integral over the boundary of the contour,
\begin{equation}
\delta I(r) = \int\!d\xi\ \det A\ \delta(R(\xi) - r) \
\delta R(\xi)\ F(z(\xi))
\;\;.
\end{equation}
Here 
\begin{equation}
\delta R(\xi) \equiv 
\sum_{i,j}
\frac{\partial R(\xi)}{\partial \xi^i}\
B^i_j(\xi) \ \frac{\partial z^j(\xi)}{\partial \lambda}\
\delta\lambda
\;\;,
\end{equation}
where $B$ is the inverse matrix to $A$,
\begin{equation}
\sum_j
A^i_j B^j_k = \delta^i_k
\;\;.
\end{equation}
We think of $A$ as producing a vector $\delta z$ from a vector $\delta \xi$, $\delta z^i = \sum_j A^i_j\, \delta \xi^j$. Then we can think of $B$ as producing a vector $\delta \xi$ from the vector $\delta z$ given by the change of $z$ under the change of deformation,
\begin{equation}
\delta \xi^i = \sum_{j}B^i_j(\xi) \ 
\frac{\partial z^j(\xi)}{\partial \lambda}\,\delta\lambda
\;\;.
\end{equation}
This justifies the name $\delta R$ for the combination
\begin{equation}
\delta R = \sum_i \frac{\partial R}{\partial \xi^i}\,
\delta \xi^i
\;\;.
\end{equation}

Given the ansatz~(\ref{eq:contour}) for $z(\xi)$, the variation $\delta z$ takes the form
\begin{equation}
\delta z^k = \frac{\mi\delta\lambda}
                  {[1 + \mi \lambda\sum_{j=1}^N \eta^j(\xi)]^2}\
\left[
\eta^k(\xi) - \xi^k \sum_{j=1}^N \eta^j(\xi)
\right]
\;\;.
\label{eq:zvariation}
\end{equation}
We build into the definition of the contour deformation the requirement that as any $\xi^k$ vanishes, the corresponding $\eta^k$ also vanishes, with $\eta^k \propto \xi^k$. Then $\delta z^k \propto \xi^k$ in this limit. Also, when $1 - \sum\xi^k \to 0$, it follows from Eq.~(\ref{eq:zvariation}) that $\sum\delta z^k \propto 1 - \sum\xi^k$. The result is that as we approach a boundary of the integration region, $R(\xi) \to 0$, the function $\delta R(\xi)$ vanishes, with 
\begin{equation}
\delta R(\xi) = R(\xi)\times h(\xi)
\;\;,
\end{equation}
where $h(\xi)$ is non-singular. Thus
\begin{equation}
\delta I(r) = r \times \int\!d\xi\ \det A\ \delta(R(\xi) - r) \
h(\xi)\ F(z(\xi))
\;\;.
\end{equation}
The factor $r$ would seem to imply that $\delta I(r) \to 0$ as $r \to 0$. However, we should be careful because $F(z(\xi))$ is singular near the boundary of the integration region. To examine this issue, we note that
\begin{equation}
I = \int_0^{1/(N+1)}\!dr\
\tilde I(r)
\;\;,
\end{equation}
where
\begin{equation}
\tilde I(r) =
\int\!d\xi\ \det A\ \delta(R(\xi) - r) \
F(z(\xi))
\;\;.
\end{equation}
Were it not for the numerator function (and the UV subtraction in the case $N=4$), the integral for $I$ would be logarithmically divergent. Generically, a one loop integral could produce two logs, so that $\tilde I(r)$ would have a singularity $\log(r)/r$ for $r \to 0$. However, the numerator factor (and UV subtraction if needed) produces an extra factor of $r$. Thus $\tilde I(r) \propto r^0 \log^K(r)$ for $r \to 0$ for some $K$. The power counting for $\delta I(r)/r$, with its extra non-singular factor $h(\xi)$, is the same. Thus $\delta I(r)$ is proportional to $r$ times possible logarithms of $r$ as $r \to 0$. 

We conclude that when we make an infinitesimal change of contour with the properties specified in this paper, the variation of the integral vanishes,
\begin{equation}
\delta I =
\lim_{r \to 0} \delta I (r) = 0
\;\;.
\end{equation}
Thus the integral on the deformed contour is the same as on the original infinitesimally deformed contour.

\section{Extra deformation}
\label{app:extradeform}

Here we return to the question of contour deformation. We study a problem that can occur with the standard deformation. We start by stating the problem rather abstractly. Let ${\cal L}$ be a subset of $\{1,2,\dots,N\}$ and let ${\cal B}$ be its complement. Suppose that
\begin{equation}
S_{ij} = 0 \hskip 1 cm {\rm for}\ i,j \in {\cal B}
\;\;.
\end{equation}
We consider the following limit. Define 
\begin{equation}
\begin{split}
\bar\xi_{\cal L} ={}& \sum_{j \in {\cal L}} \xi^j
\end{split}
\end{equation}
and let
\begin{equation}
\begin{split}
\xi^j ={}& \bar\xi_{\cal L}\,\hat\xi^j_{\cal L} 
\hskip 2.9 cm {\rm for}\ j \in {\cal L}
\;\;,
\\
\xi^j ={}& (1 - \xi^0 - \bar\xi_{\cal L})\,
\hat\xi^j_{\cal B}  \hskip 1.0 cm {\rm for}\ j \in {\cal B}
\;\;,
\label{eq:limitdef}
\end{split}
\end{equation}
so that
\begin{equation}
\sum_{j \in {\cal L}} \hat\xi^j
=\sum_{j \in {\cal B}} \hat\xi^j
=1
\;\;.
\end{equation}
Then we consider the limit $\bar\xi_{\cal L} \to 0$ with $\xi^0 = 0$. Thus the $\xi^i$ for $i \in {\cal B}$ are big and the $\xi^i$ for $i \in {\cal L}$ are little.

In the limit $\bar\xi_{\cal L} \to 0$, ${\cal S}(\xi)$ becomes
\begin{equation}
{\cal S}(\xi) = 
\sum_{i \in {\cal L}} \xi^i \sum_{j \in {\cal B}} S_{ij} \xi^j
+ \frac{1}{2} \sum_{i,j \in {\cal L}} \xi^i \xi^j S_{ij}
= 
\bar\xi_{\cal L}
\sum_{i \in {\cal L}} \hat\xi^i_{\cal L} 
\tilde w_i(\hat\xi_{\cal B})
+{\cal O}(\bar\xi_{\cal L}^2)
\;\;,
\end{equation}
where
\begin{equation}
\tilde w_i(\hat\xi_{\cal B}) = \sum_{j \in {\cal B}} S_{ij} \hat\xi^j_{\cal B}
\;\;.
\label{eq:tildewdef}
\end{equation}
If we adopt the standard contour definition from Eq.~(\ref{eq:etadef}), we have
\begin{equation}
{\cal S}(\xi + \mi \eta) =
\bar\xi_{\cal L}
\sum_{i \in {\cal L}} \hat\xi^i_{\cal L} 
\tilde w_i(\hat\xi_{\cal B})
+\mi \,\frac{\lambda}{m^2}\, \bar\xi_{\cal L}
\sum_{i \in {\cal L}} \hat\xi^i_{\cal L}
[\tilde w_i(\hat\xi_{\cal B})]^2
+ {\cal O}(\bar\xi_{\cal L}^2)
\;\;.
\end{equation}
We see that the surface $\bar \xi_{\cal L} = 0$ with $\xi^0 = 0$ is a singular surface of the integrand. In fact, it is a pinch singular surface. For a generic point $\hat \xi$, ${\cal S}(\xi + \mi \eta)$ vanishes linearly with $\bar \xi_{\cal L}$ as $\bar \xi_{\cal L} \to 0$.

This generic behavior is fine from a numerical point of view. However, we would like to avoid having ${\cal S}(\xi + \mi \eta)$ vanish faster than linearly as $\bar \xi_{\cal L} \to 0$. The real part of ${\cal S}(\xi + \mi \eta)$ can easily vanish quadratically as $\bar \xi_{\cal L} \to 0$. The components $S_{ij}$ can have either sign, so that for some points $\hat \xi$ the particular linear combination $\sum_{i \in {\cal L}} \hat\xi^i_{\cal L}  \tilde w_i(\hat\xi_{\cal B})$ can vanish. It is harder for the linear contribution to the imaginary part of ${\cal S}(\xi + \mi \eta)$ to vanish. However, if the set ${\cal B}$ has more than one element, it is possible for $\tilde w_i(\hat\xi_{\cal B})$ to vanish for some particular index $i = I$ at some particular value of $\hat\xi_{\cal B}$. Then if all of the $\hat\xi^i_{\cal L}$ vanish except for $i = I$, we will have
\begin{equation}
\sum_{i \in {\cal L}} \hat\xi^i_{\cal L}
[\tilde w_i(\hat\xi_{\cal B})]^2
= \hat\xi^{I}_{\cal L}
[\tilde w_{I}(\hat\xi_{\cal L})]^2
= 0
\;\;.
\end{equation}
For this choice of the $\hat \xi^i$, we will have $\Lambda^2(\xi + \mi \eta) = {\cal O}(\bar\xi_{\cal L}^2)$ if we take the standard deformation. One might think that having an esoteric integration region in which the integrand is extra singular is not a problem. However, in a numerical integration it is a problem. One possibility is to put extra integration points in the region of extra singularity, but a more attractive possibility is to fix the contour deformation so as to better keep the integration contour away from the singularity. At the same time, we need to avoid letting the jacobian $\det\!\left({dz}/{d\xi}\right)$ in Eq.~(\ref{eq:lxspacedeformed}) become singular. This is the strategy we will pursue.

Until now, we have followed a rather abstract formulation of the problem for the reason that the same abstract problem occurs in several ways in the subtraction terms defined in Ref.~\cite{NSsubtractions} to take care of infrared divergent graphs. In this paper, however, we are concerned with infrared finite graphs representing photon scattering with a massless electron loop. The problem is associated with the region in the original loop integral in which lines $n$ and $n+1$ are nearly collinear.  With massless kinematics, $S_{nn} = S_{n+1,n+1} = S_{n,n+1} = 0$. Thus the matrix $S_{ij}$ has the special form with ${\cal B} = \{n,n+1\}$ and ${\cal L}$ consisting of all index values $i \in \{1,\dots,N\}$ other than $n$ and $n+1$.

We will seek a supplementary deformation $\tilde\eta^n(\xi)$ and $\tilde\eta^{n+1}(\xi)$ that we can add to the standard deformation. In the following, we consider $n$ to be any fixed index value in the range $1 \le n \le N$. There will be an analogous deformation for any $n$. For reasons that will become apparent, we want to have just one of these added deformations $\tilde \eta(\xi)$ for any value of $\xi$. For this reason, we will arrange that 
$\tilde\eta^n(\xi)$ and $\tilde\eta^{n+1}(\xi)$ are nonzero only in the region
\begin{equation}
\label{eq:Rndef}
{\cal R}_n:\quad
\xi^n + \xi^{n+1} > {\textstyle\frac{1}{2}},\
\xi^n > \xi^{n+2},\
\xi^{n+1} > \xi^{n-1}
\;\;.
\end{equation}
It is easy to verify that the various regions ${\cal R}_n$ are non-overlapping.

Given that there is an added deformation $\tilde\eta^n(\xi)$ and $\tilde\eta^{n+1}(\xi)$, there is an added contribution to ${\rm Im}\,{\cal S}$ that has the form
\begin{equation}
\begin{split}
\Delta\,{\rm Im}\,{\cal S}(\xi + \mi \eta) ={}& 
\sum_{i \in {\cal L}}\xi^i
\left\{
\tilde\eta^{n}  S_{n,i}
+\tilde\eta^{n+1}  S_{n+1,i} 
\right\}
\;\;.
\label{eq:DeltaLambda}
\end{split}
\end{equation}
When we include the standard contour definition from Eq.~(\ref{eq:etadef}), the total ${\rm Im}\,{\cal S}$ is
\begin{equation}
\begin{split}
{\rm Im}\,{\cal S}(\xi + \mi \eta) ={}& 
\sum_{i \in {\cal L}} \xi^i 
\left\{
\tilde\eta^{n}  S_{n,i}
+\tilde\eta^{n+1}  S_{n+1,i}
+(\lambda/m^2)[w_i(\xi)]^2
\right\}
\\ &
+\sum_{i\in {\cal B}} 
(\lambda/m^2) \xi^i [w_i(\xi)]^2
\;\;.
\label{eq:ImLambdamod1}
\end{split}
\end{equation}
We take the extra deformations to be of the form $\pm (\lambda/m^2)\, g(\xi)$: 
\begin{equation}
\begin{split}
\label{eq:tildeetadef}
\tilde\eta^n ={}& 
(\lambda/m^2)\, g(\xi) \;,
\\
\tilde\eta^{n+1} ={}& 
 - (\lambda/m^2)\, g(\xi) \;.
\end{split}
\end{equation}
Here $g(\xi)$ is a function to be defined. 

With the definition (\ref{eq:tildeetadef}) for the added deformation together with the standard deformation, we have
\begin{equation}
\begin{split}
{\rm Im}\,{\cal S}(\xi + \mi \eta) ={}& 
(\lambda/m^2)
\sum_{i \in {\cal L}} \xi^i 
\left\{
(S_{n,i} - S_{n+1,i})g(\xi)
+[w_i(\xi)]^2
\right\}
\\ &+
\sum_{i\in {\cal L}} 
(\lambda/m^2) \xi^i [w_i(\xi)]^2
\;\;.
\label{eq:ImLambdamod2}
\end{split}
\end{equation}
The second term here is always positive but vanishes quadratically with $\bar\xi_{\cal L}$ in the limit $\bar\xi_{\cal L} \to 0$, so it is too small to help us in this limit. In the first term, the parts proportional to  $[w_j(\xi)]^2$ are always positive and for typical values of $\xi^n$ and $\xi^{n+1}$ vanish linearly with $\bar\xi_{\cal L}$. However, $w_i(\xi)$ for some value of $i$ can vanish in this limit for a particular value of $\xi^n$ and $\xi^{n+1}$. This is the reason for adding the new deformation specified by $g(\xi)$.

We need to ensure that
\begin{equation}
\label{eq:goal}
(S_{n,i} - S_{n+1,i})g(\xi)
+[w_i(\xi)]^2
> 0
\end{equation}
for all $i \in {\cal L}$ and for all $\xi$. Here we really want a ``$>$'' relation and not just a ``$\ge$'' relation, which could be satisfied with $g(\xi) = 0$.

This goal can be accomplished by a straightforward construction. First, we need some notation indicating certain sets of indices. Let ${\cal L}_{+}$ be set of indices in ${\cal L}$ such that $S_{n+1,i} > S_{n,i}$ and let ${\cal L}_{-}$ be set of indices in ${\cal L}$ such that $S_{n,i} > S_{n+1,i}$. The union of ${\cal L}_{+}$ and ${\cal L}_{-}$ is all of ${\cal L}$. (Here we assume that the external momenta do not lie on the surface $S_{n+1,i} = S_{n,i}$ for any $i$ in $\cal L$.)

Define
\begin{equation}
\begin{split}
g_+(\xi) ={}& \min_{i \in {\cal L}_+} \frac{[w_i(\xi)]^2}{S_{n+1,i} - S_{n,i}}
\;\;,
\\
g_-(\xi) ={}& \min_{i \in {\cal L}_-} \frac{[w_i(\xi)]^2}{S_{n,i} - S_{n+1,i}}
\;\;.
\end{split}
\end{equation}
Then Eq.~(\ref{eq:goal}) requires that
\begin{equation}
\label{eq:goal2}
- g_-(\xi) < g(\xi) < g_+(\xi)
\;\;.
\end{equation}
Of particular interest is the requirement for a special point for which $w_i(\xi) = 0$. If $i \in {\cal L}_-$, then $g_-(\xi) = 0$ at this point and the requirement is that $g(\xi)$ be positive (but not too positive). If $i \in {\cal L}_+$, then $g_+(\xi) = 0$ at this point and the requirement is that $g(\xi)$ be negative (but not too negative).

These restrictions are not very restrictive in the case that either $g_+(\xi)$ or $g_-(\xi)$ is large or in the case that one of the sets ${\cal L}_\pm$ is empty (in which case we interpret the corresponding $g_\pm$ to be infinite). In order to ensure that the deformation not be too large, we can also impose
\begin{equation}
\label{eq:goal3}
- \lambda_g < (\lambda/m^2)\,g(\xi) < \lambda_g
\;\;.
\end{equation}
where $\lambda_g$ is a parameter that could be chosen to be $\lambda$.
Defining
\begin{equation}
\tilde g_\pm(\xi) = \min[g_\pm(\xi),m^2 \lambda_g/\lambda]
\;\;,
\end{equation}
our requirement is
\begin{equation}
\label{eq:goal4}
- \tilde g_-(\xi) < g(\xi) < \tilde g_+(\xi)
\;\;.
\end{equation}
It is easy to satisfy Eq.~(\ref{eq:goal4}). We set
\begin{equation}
\label{eq:gdef}
g(\xi) = H(\xi)\, C_g
\big[ \tilde g_+(\xi) - \tilde g_-(\xi) \big]
\;\;,
\end{equation}
where $C_g$ is a parameter in the range $0 < C_g < 1$ (possibly 1/2) and
\begin{equation}
\begin{split}
\label{eq:Hdef}
H(\xi) ={}& 
(2\xi^n + 2\xi^{n+1} - 1)\
4(\xi^n - \xi^{n+2})(\xi^{n+1} - \xi^{n-1})
\\& \times
\theta(\xi^n + \xi^{n+1} > {\textstyle\frac{1}{2}})\
\theta(\xi^n > \xi^{n+2})\
\theta(\xi^{n+1} > \xi^{n-1})
\;\;.
\end{split}
\end{equation}
The purpose of $H(\xi)$ is to restrict the range of $\xi$ for which $g(\xi) \ne 0$ to the desired region ${\cal R}_n$, Eq.~(\ref{eq:Rndef}). There is also a factor that becomes $4 \xi^n \xi^{n+1}$ in the limit $\bar \xi_{\cal L} \to 0$. This factor turns off the deformation as $\xi^n \to 0$ or $\xi^{n+1} \to 0$. Notice that
\begin{equation}
0 \le H(\xi) \le 1
\;\;.
\end{equation}
With the use of this property, it is evident that the definition (\ref{eq:gdef}) satisfies Eq.~(\ref{eq:goal4}).

Suppose that that there is an index $i \in {\cal L}_+$ and an index $j \in {\cal L}_-$ such that 
\begin{equation}
\begin{split}
S_{i,n+1} >{}& 0 \hskip 1 cm S_{i,n} < 0
\;\;,
\\
S_{j,n} >{}& 0 \hskip 1 cm S_{j,n+1} < 0
\;\;,
\end{split}
\end{equation}
and that
\begin{equation}
S_{i,n}S_{j,n+1} - S_{j,n}S_{i,n+1} \ll s^2
\;\;.
\end{equation}
Then there is approximately an effective contour pinch, in the sense that both $g^+(\xi)$ and $g^-(\xi)$ are close to vanishing when all of the  $\xi^k$ for $k \in {\cal L}$ are very small and
\begin{equation}
\begin{split}
\xi^n ={}& \frac{S_{i,n+1} - S_{j,n+1}}
{S_{i,n+1} - S_{i,n} + S_{j,n} - S_{j,n+1}}
\;\;,
\\
\xi^{n+1} ={}& \frac{ S_{j,n} - S_{i,n}}
{S_{i,n+1} - S_{i,n} + S_{j,n} - S_{j,n+1}}
\;\;.
\end{split}
\end{equation}
The contour is pinched already along the whole collinear singularity line $\xi^k = 0$ for $k \in {\cal L}$, but this is an extra pinch that prevents the deformation of this section from being effective. For this reason, one should put extra integration points in the region near this point.

It is instructive to examine the functions $w_i(\xi)$ for $i \in {\cal L}$ in the limit that $\xi^0$ and all of the $\xi^j$ for $j \in {\cal L}$ vanish (and assuming massless kinematics). Then the only two $x^i$ that are non-zero are $\xi^n$ and $\xi^{n+1} = 1 - \xi^n$. Following the notation of Eq.~(\ref{eq:tildewdef}), we can call the limiting function $\tilde w_i(\xi^n)$. We would like to know for what value of $\xi^n$ (if any) this function vanishes. We have
\begin{equation}
\tilde w_i(\xi^n) = S_{in}\xi^n + S_{i,n+1}(1 - \xi^n)
\;\;.
\end{equation}
Evidently $\tilde w_i(\xi^n)$ will vanish for some $\xi^n$ in the range $0 < \xi^n < 1$ if and only if $S_{in}$ and $S_{i,n+1}$ are non-zero and have opposite signs.

Thus we need to know something about the signs of $S_{in}$ and $S_{i,n+1}$. First, we note that if $i = n-1$ then $S_{in} = 0$. Furthermore, if all of the particles $i,i+1,\dots,n-1$ are final state particles, then 
\begin{equation}
S_{in} = \left(
\sum_{j = i}^{n-1} P_j
\right)^2 > 0
\;\;.
\end{equation}
If two of the particles $i,i+1,\dots,n-1$ are the two initial state particles, then
\begin{equation}
S_{in} = \left(
\sum_{j = n}^{i-1} P_j
\right)^2 > 0
\;\;.
\end{equation}
If exactly one of the particles $i,i+1,\dots,n-1$ is an initial state particle, then $S_{in} < 0$. The proof of this amounts to showing that if a massless particle $A$ turns into a massive particle $A'$ by exchanging a momentum $Q$ and a massless particle $B$ turns into a massive particle $B'$ by absorbing the momentum $Q$, then $Q^2 < 0$. We omit the details. 

Given these results and the analogous results for $S_{i,n+1}$ we can conclude that $S_{in}$ and $S_{i,n+1}$ are non-zero and have opposite signs when $i$ is neither of $n-1$ or $n+2$ and external photon $n$ is an incoming particle.

Supposing that photon $n$ is an incoming particle, then the propagator index $i$ is in ${\cal L}_-$, with $S_{in} > 0$ and $S_{i,n+1} < 0$ if all of the external particles $i,i+1,\dots,n-1$ are final state particles. If, on the other hand, one of them is the other initial state particle, then $S_{in} < 0$ and $S_{i,n+1} > 0$ and $i$ is in ${\cal L}_+$.

As we have seen, when photon $n$ is an incoming particle, the zero of $\tilde w_i(\xi^n)$, 
\begin{equation}
\xi^n_{(i)} = \frac{S_{i,n+1}}{S_{i,n+1} - S_{i,n}}
\;\;,
\end{equation}
lies in the integration range, $0 < \xi^n < 1$. In this case, we can say more about the location of this zero. Let the index of the other incoming photon be $n'$. Define
\begin{equation}
\tau = - 2 P_{n'} \cdot \sum_{j = n+1}^{n'-1} P_k /s
\;\;.
\end{equation}
Then if the index $i$ is in the range $n'+1,\dots,n-1$, we have $S_{i,n} > 0$ and $S_{i,n+1} < 0$ so $i \in {\cal L}_-$. Then one can show that
\begin{equation}
\xi^n_{(i)} > \tau 
\;\;.
\label{tauinequality1}
\end{equation}
On the other hand, if the index $i$ is in the range $n+2,\dots,n'$, we have $S_{i,n} < 0$ and $S_{i,n+1} > 0$ so $i \in {\cal L}_+$. Then one can show that
\begin{equation}
\xi^n_{(i)} < \tau 
\;\;.
\label{tauinequality2}
\end{equation}

To prove Eq.~(\ref{tauinequality2}), write each outgoing momentum in the form
\begin{equation}
P_i = -a_i P_n - b_i P_{n'} + P_i^T
\end{equation}
where $P_i^T \cdot P_n = P_i^T \cdot P_{n'} = 0$ and $0 < a_i < 1$ and $0 < b_i < 1$. Then for $i$ in the range $n+2,\dots,n'$ we have
\begin{equation}
S_{i,n+1} = \left(\sum_{j = n+1}^{i-1} P_j\right)^2
= \left(\sum_{j = n+1}^{i-1} a_j\right)\left(\sum_{j = n+1}^{i-1} b_j\right)s
+ \left(\sum_{j = n+1}^{i-1} P_j^T\right)^2
\;\;.
\end{equation}
For $S_{i,n}$ we add one more particle $n$ with $a_n = -1$, $b_n = 0$, and no $P_n^T$. Thus
\begin{equation}
S_{i,n} = \left(\sum_{j = n}^{i-1} P_j\right)^2
= -\left(1 - \sum_{j = n+1}^{i-1} a_j\right)
\left(\sum_{j = n+1}^{i-1} b_j\right)s
+ \left(\sum_{j = n+1}^{i-1} P_j^T\right)^2
\;\;.
\end{equation}
Then
\begin{equation}
\xi^n_{(i)} = \frac{\left(\sum_{j = i}^{n-1} a_j\right)
\left(\sum_{j = i}^{n-1} b_j\right)s
+ \left(\sum_{j = i}^{n-1} P_j^T\right)^2}
{\left(\sum_{j = i}^{n-1} b_j\right)s}
\;\;,
\end{equation}
The $(P^T)^2$ term in the numerator is negative. Thus
\begin{equation}
\xi^n_{(i)} < \sum_{j = i}^{n-1} a_j
< \sum_{j = n'}^{n-1} a_j 
=  \tau
\;\;.
\end{equation}
The proof of Eq.~(\ref{tauinequality1}) is similar.

Thus in the limit that all of the $\xi^j$ for $j \in {\cal L}$ are very small, the qualitative nature of the deformation constructed here is quite simple. We deform $\xi^n$ into the upper half plane for $\xi > \tau$ and into the lower half plane for $\xi < \tau$.

\section{Double parton scattering singularity}
\label{sec:appendixdps}

As discussed briefly in Sec.~\ref{sec:dps}, a pinch singular point corresponding to double parton scattering can be present if a special condition holds for the external momenta. This singularity is illustrated in Fig.~\ref{fig:dps}. Imagine that incoming parton with index $A$ carries momentum $-P_A$ such that $P_A^2 = 0$ and that parton $A$ splits into two collinear partons with labels $A$ and $A+1$. That is $-K_A(\xi) = -(1-x_A) P_A$ and $K_{A+1}(\xi) = -x_A P_A$. Imagine also that incoming parton with index $B$ carries momentum $-P_B$ such that $P_B^2 = 0$ and that parton $B$ splits into two collinear partons with labels $B$ and $B+1$. That is $-K_B(\xi) = (1-x_B) P_B$ and $K_{B+1}(\xi) = - x_B  P_B$. (Here the $x$'s are momentum fractions, not Feynman parameters.) Partons $A+1$ and $B$ could meet and produce a group of final state partons with labels $i$ in a set ${\cal A} = \{A+1,\dots,B-1\}$.  Partons $B+1$ and $A$ could meet and produce a group of final state partons with labels $i$ in a set ${\cal B}= \{B+1,\dots,A-1\}$. Thus
\begin{equation}
\begin{split}
\sum_{i \in {\cal A}} P_i ={}& - x_A P_A - (1-x_B)P_B
\;\;, 
\\
\sum_{i \in {\cal B}} P_i ={}& - (1 - x_A) P_A -  x_B P_B
\;\;.
\label{eq:dps1}
\end{split}
\end{equation}
It is convenient to write Eq.~(\ref{eq:dps1}) in terms of the internal line momenta $Q_i$ using Eq.~(\ref{eq:Qndef}). We note immediately that $\sum_{i \in {\cal A}} P_i = Q_{B} - Q_{A+1}$ and $\sum_{i \in {\cal B}} P_i = Q_{A} - Q_{B+1}$ are timelike vectors. Thus
\begin{equation}
\begin{split}
S_{A+1,B} >{}& 0 \;\;,
\\
S_{A,B+1} >{}& 0 \;\;.
\end{split}
\end{equation}
Notice that for this kind of singularity to occur, we need at least two external lines in set ${\cal A}$ and two in set ${\cal B}$. Thus we need at least four outgoing external particles. Thus we need $N \ge 6$.

Given the external momenta, the momentum fractions $x_A$ and $x_B$ are determined. When rewritten in terms of the $Q_i$, the first of Eq.~(\ref{eq:dps1}) reads
\begin{equation}
Q_{A+1} - Q_{B+1} = x_A (Q_{A+1} - Q_A) - x_B (Q_{B+1} - Q_B)
\;\;,
\label{eq:dps2}
\end{equation}
while the second equation in (\ref{eq:dps1}) is just the negative of this. Take the inner product of this with $(Q_{B+1} - Q_B)$ and use
\begin{equation}
2(Q_{A+1} - Q_A)\cdot (Q_{B+1} - Q_B) = \overline S
\;\;,
\end{equation}
where
\begin{equation}
\overline S \equiv S_{A,B+1} + S_{A+1,B} - S_{A,B} - S_{A+1,B+1}
\;\;.
\label{eq:overlineSdef}
\end{equation}
Also note that $(Q_{B+1} - Q_B)^2 = 0$ and
\begin{equation}
2(Q_{A+1} - Q_{B+1})\cdot (Q_{B+1} - Q_B)
= S_{A+1,B} - S_{A+1,B+1} \;\;.
\end{equation}
These relations give
\begin{equation}
x_A = \frac{S_{A+1,B} - S_{A+1,B+1}}
{\overline S}
\;\;.
\end{equation}
We similarly derive
\begin{equation}
x_B = \frac{S_{A,B+1} - S_{A+1,B+1}}
{\overline S}
\;\;.
\end{equation}

The kinematic conditions require that
\begin{equation}
\label{eq:doublescatteringcondition}
\sum_{i \in {\cal A}} P_i^{\rm T} = 0
\;\;,
\end{equation}
where $P_i^{\rm T}$ is the part of $P_i$ transverse to $P_A$ and $P_B$. (In this frame the sum of the transverse momenta of all of the final state particles vanishes, so the sum of the $P_i^{\rm T}$ for the particles in set ${\cal B}$ also vanishes if the sum for set ${\cal A}$ vanishes.) The condition for this to happen is obtained by squaring both sides of Eq.~(\ref{eq:dps2}) and inserting the solutions for $x_A$ and $x_B$. This gives 
\begin{equation}
\label{eq:detiszero}
S_{A+1,B}\, S_{A,B+1} - S_{A,B}\, S_{A+1,B+1}  = 0
\;\;.
\end{equation}
That is,
\begin{equation}
\det
\left(
\begin{matrix}
S_{A,B} & S_{A+1,B} \\
S_{A,B+1} & S_{A+1,B+1}
\end{matrix}
\right)
= 0
\;\;.
\end{equation}

Recall that in order to have a double parton scattering singularity, $S_{A+1,B}>0$ and $S_{A,B+1}>0$. The determinant condition then implies that $S_{A,B}$ and $S_{A+1,B+1}$ have the same sign. In fact, this sign must be negative. To see this, one may note that, because of Eq.~(\ref{eq:detiszero}), two alternative expressions for $x_A$ are also valid:
\begin{equation}
x_A = \frac{S_{A+1,B}}{S_{A+1,B} - S_{A,B}}
=
\frac{-S_{A+1,B+1}}{S_{A,B+1} - S_{A+1,B+1}}
=
\frac{S_{A+1,B}-S_{A+1,B+1}}
{\overline S}
\;\;.
\end{equation}
Using the first of these, we see that $x_A > 0$ implies that $S_{A+1,B} - S_{A,B} > 0$. But then $x_A < 1$ implies that $S_{A,B} < 0$. We conclude that for a double parton scattering singularity, $S_{A+1,B}>0$ and $S_{A+1,B}>0$, $S_{A,B} < 0$ and $S_{A+1,B+1} < 0$.

What does this mean in terms of solving Eq.~(\ref{eq:pinchcondition1})? We demand that Eq.~(\ref{eq:pinchcondition1}) hold for nonzero $\xi^A$, $\xi^{A+1}$, $\xi^B$ and $\xi^{B+1}$ with all of the other $\xi^i = 0$. Thus we need $w_B = w_{B+1} = 0$, or
\begin{equation}
\left(
\begin{matrix}
S_{A,B} & S_{A+1,B} \\
S_{A,B+1} & S_{A+1,B+1}
\end{matrix}
\right)
\left(
\begin{matrix}
\xi^A  \\
\xi^{A+1} 
\end{matrix}
\right)
= 0
\;\;.
\end{equation}
Similarly we need $w_A = w_{A+1} = 0$, or
\begin{equation}
\left(
\begin{matrix}
S_{A,B} & S_{A,B+1} \\
S_{A+1,B} & S_{A+1,B+1}
\end{matrix}
\right)
\left(
\begin{matrix}
\xi^B  \\
\xi^{B+1} 
\end{matrix}
\right)
= 0
\;\;.
\end{equation}
One can solve these if the determinant of the matrix is zero, that is if 
Eq.~(\ref{eq:detiszero}) holds. If it does, the solution with $\xi^A + \xi^{A+1} + \xi^B + \xi^{B+1} = 1$ is
\begin{equation}
\begin{split}
\xi^A ={}& f_A\, \bar x \;\;,
\\
\xi^{A+1} ={}& (1-f_A)\, \bar x \;\;,
\\
\xi^B ={}& f_B\, (1- \bar x) \;\;,
\\
\xi^{B+1} ={}& (1-f_B)\, (1 - \bar x )\;\;,
\label{eq:xifordps}
\end{split}
\end{equation}
where
\begin{equation}
\begin{split}
f_A ={}& \frac{S_{A+1,B}}{S_{A+1,B} - S_{A,B}}
=
\frac{-S_{A+1,B+1}}{S_{A,B+1} - S_{A+1,B+1}}
=
\frac{S_{A+1,B}-S_{A+1,B+1}}
{\overline S}
\;\;,
\\
f_B ={}& \frac{S_{A,B+1}}{S_{A,B+1} - S_{A,B}}
=
\frac{- S_{A+1,B+1}}{S_{A+1,B} - S_{A+1,B+1}}
=
\frac{S_{A,B+1} - S_{A+1,B+1}}
{\overline S} \;\;.
\end{split}
\end{equation}
That is, $f_A = x_A$ and $f_B = x_B$. In order for the pinch singularity to be inside the integration region, $\xi^A$, $\xi^{A+1}$, $\xi^B$ and $\xi^{B+1}$ need to be positive. Thus we need to choose $\bar x$ in the range
\begin{equation}
0 < \bar x < 1
\;\;.
\end{equation}

It is of interest to work out the momenta $K_i(\xi)$, Eq.~(\ref{eq:Kndef}), when the external momenta obey the condition~(\ref{eq:dps2}) for a double parton scattering singularity and the Feynman parameters $\xi$ are given by Eq.~(\ref{eq:xifordps}). One finds
\begin{equation}
\begin{split}
K_A(\xi) ={}& (1-f_A) P_A \;\;,
\\
K_{A+1}(\xi) ={}&  -f_A  P_A \;\;,
\\
K_B(\xi) ={}& (1-f_B) P_B \;\;,
\\
K_{B+1}(\xi) ={}&  -f_B  P_B \;\;.
\end{split}
\end{equation}
These are, of course, the relations we started with.

We learn that if the determinant condition (\ref{eq:detiszero}) and certain sign conditions hold, $\Lambda^2(\xi)$ has a pinch singularity along a line that runs through the middle of the integration region. Now, the pinch singularity conditions hold only for certain special choices of the external momenta. However, one can easily be near to having a pinch singularity. For this reason, in a numerical program, one should check for each graph if $|S_{A+1,B}\, S_{A,B+1} - S_{A,B}\, S_{A+1,B+1}| \ll \overline S^2$, with the required sign conditions, for some choice of indices $A$ and $B$. In that event, one should put a high density of integration points near the ``almost'' singular line.


\end{document}